\documentclass[pre,showpacs,twocolumn,amsmath,amssymb,superscriptaddress,floatfix]{revtex4}
\usepackage{graphicx}
\usepackage{dcolumn}
\usepackage{bm}
\usepackage{latexsym}

\begin{document}

\title{Phase Transitions and Spatio-Temporal Fluctuations in Stochastic 
	Lattice Lotka--Volterra Models}

\author{Mauro Mobilia} \email{mauro.mobilia@physik.lmu.de}
\affiliation{Arnold Sommerfeld Center for Theoretical Physics and 
        Center for NanoScience, Department of Physics, 
        Ludwig-Maximilians-Universit\"at M\"unchen, D-80333 Munich, Germany}
\affiliation{Department of Physics and 
     	Center for Stochastic Processes in Science and Engineering, 
     	Virginia Polytechnic Institute and State University,  
     	Blacksburg, Virginia 24061-0435, U.S.A.}	
\author{Ivan T. Georgiev} \email{igeorgiev@ifltd.com}
\affiliation{Integrated Finance Limited, 630 Fifth Avenue, Suite 450, 
        New York, NY 10111, U.S.A.}
\affiliation{Department of Physics and 
        Center for Stochastic Processes in Science and Engineering, 
        Virginia Polytechnic Institute and State University,  
        Blacksburg, Virginia 24061-0435, U.S.A.}
\author{Uwe C. T\"auber}  \email{tauber@vt.edu}
\affiliation{Department of Physics and 
        Center for Stochastic Processes in Science and Engineering, 
        Virginia Polytechnic Institute and State University,  
        Blacksburg, Virginia 24061-0435, U.S.A.}     

\begin{abstract}
We study the general properties of stochastic two-species models for
predator-prey competition and coexistence with Lotka--Volterra type
interactions defined on a $d$-dimensional lattice. 
Introducing spatial degrees of freedom and allowing for stochastic 
fluctuations generically invalidates the classical, deterministic mean-field 
picture.
Already within mean-field theory, however, spatial constraints, modeling 
locally limited resources, lead to the emergence of a continuous 
active-to-absorbing state phase transition.
Field-theoretic arguments, supported by Monte Carlo simulation results, 
indicate that this transition, which represents an extinction threshold for 
the predator population, is governed by the directed percolation universality 
class. 
In the active state, where predators and prey coexist, the classical center
singularities with associated population cycles are replaced by either nodes 
or foci.
In the vicinity of the stable nodes, the system is characterized by 
essentially stationary localized clusters of predators in a sea of prey.
Near the stable foci, however, the stochastic lattice Lotka--Volterra system 
displays complex, correlated spatio-temporal patterns of competing activity 
fronts. 
Correspondingly, the population densities in our numerical simulations turn 
out to oscillate irregularly in time, with amplitudes that tend to zero in the 
thermodynamic limit.
Yet in finite systems these oscillatory fluctuations are quite persistent, and
their features are determined by the intrinsic interaction rates rather than 
the initial conditions. 
We emphasize the robustness of this scenario with respect to various model 
perturbations.
\end{abstract}


\date{\today}

\maketitle

\section{Introduction}

Since Lotka and Volterra's seminal and pioneering works \cite{Lotka,Volterra},
many decades ago, modeling of interacting, competing species has received 
considerable attention in the fields of biology, ecology, mathematics 
\cite{Haken,Neal,May1,Maynard,Murray,Kolmogorov,Montroll,Picard,Sigmund,
Leonard,Durrett,data,spatial}, and, more recently, in the physics literature 
as well \cite{Tome,Lip,TiborDroz,McKane,Dunbar,Provata,Matsuda,cyclic,Shnerb,
Boccara,Albano,Droz}.
In their remarkably simple deterministic model, Lotka and Volterra considered 
two coupled nonlinear differential equations that mimic the temporal evolution
of a two-species system of competing predator and prey populations.
They demonstrated that coexistence of both species was not only possible but
inevitable in their model.
Moreover, similar to observations in real populations, both predator and prey
densities in this deterministic system display regular oscillations in time, 
with both the amplitude and the period determined by the prescribed initial 
conditions (only near the center fixed point associated with the coexistence 
of the two populations is the oscillation frequency solely given in terms of 
the intrinsic interaction rates, see Sec.~II.A below).
However, despite the undisputed mathematical elegance of these results, and
its consequent ubiquity in textbooks \cite{Haken,Neal,May1,Maynard,Murray}, 
the original Lotka--Volterra model (LVM) is often severely criticized on the 
grounds of being biologically too simplistic and therefore unrealistic 
\cite{Neal}, and mathematically unstable with respect to model modifications 
\cite{Murray}.

In this paper, we aim at drawing a comprehensive, detailed picture of the 
{\em stochastic} dynamics, defined on a $d$-dimensional lattice, of two 
competing populations with Lotka--Volterra type predation interaction. 
The systems under consideration are `individual-based' lattice models, where 
each lattice site can be occupied by a given (finite) number of particles.
We shall formulate the {\em stochastic lattice Lotka--Volterra model} (SLLVM)
in the natural language of a reaction--diffusion lattice gas model, i.e., in 
terms of appropriate stochastic particle hopping and creation and annihilation
processes defined on a lattice, and will here investigate them by means of 
various methods of the theory of nonequilibrium statistical mechanics, 
including mean-field approximations, Monte Carlo computer simulations,  
field-theoretic representations and renormalization group arguments.
With these techniques, we are thus able to consider and discuss the role of 
spatial constraints, spatio-temporal fluctuations and correlations, and 
finite-size effects.
We shall argue that although the criticisms against the classical LVM 
definitely pertain to the original deterministic rate equations, introducing 
spatial degrees of freedom and allowing for stochasticity \cite{Durrett} 
actually renders the corresponding two-species reaction system considerably 
richer, definitely more interesting, and perhaps even more realistic.
In addition, in stark contrast with the deterministic LVM, the SLLVM scenario 
turns out to be remarkably robust with respect to model modifications, and 
thus appears to provide a quite generic picture of two-species 
predator-prey interactions.

In recent years, population dynamics has received considerable attention from 
the statistical physics community. 
In particular, a variety of so-called `individual-based', or stochastic, 
lattice predator-prey models have been investigated, e.g., in
Refs.~\cite{Tome,Lip,TiborDroz,McKane,Dunbar,Provata,Matsuda,cyclic}, 
employing largely mean-field-type approaches (including refined versions, such
as the pair approximation of Refs.~\cite{Tome,Matsuda}) and Monte Carlo 
simulations, mostly in two dimensions. 
Among the main issues addressed by these papers are the phase diagrams of 
these stochastic lattice Lotka--Volterra systems (see, e.g., 
Refs.~\cite{Tome,Lip,TiborDroz,Matsuda,Boccara,Albano,Droz}), the critical 
properties near the predator extinction threshold, typically argued to be 
governed by the scaling exponents of the directed percolation (DP) universality
class \cite{Tome,Lip,TiborDroz,Boccara,Albano}, and the presence or absence of 
(stochastic) oscillations, whose amplitudes were reported in 
Refs.~\cite{Provata,Lip,TiborDroz} to (globally) decrease as the system size 
was increased.

Since it would be impossible (and well beyond our scope here) to provide a 
complete survey of the numerous contributions of statistical physicists to the
fascinating field of population dynamics, we choose, for the sake of clarity, 
to briefly discuss more specifically some work on lattice predator--prey 
models that we have found to be particularly relevant for the issues 
considered in this article.
In Refs.~\cite{Lip,TiborDroz}, the authors considered various two-species
four-state models (in the absence of diffusion, each lattice site is either
empty, occupied by a single predator or prey, or by both a predator and prey) 
and noticed that both extinction and coexistence of the two populations are 
possible.
They found that there exists a sharp continuous transition between the 
predator extinction phase and the active predator--prey coexistence phase.
Numerical studies of the (static) critical properties near the predator 
extinction threshold (mainly in one and two dimensions) revealed that its 
critical exponents are consistent with those of DP \cite{DP,DPreviews} (see 
below). 
In addition, in Refs.~\cite{Lip,TiborDroz} the oscillatory behavior displayed 
by the densities of the coexisting populations in some region of the active 
parameter space was studied as well. 
It was reported that in one and two dimensions the population densities showed
characteristic erratic oscillations whose amplitude vanishes in the 
thermodynamic limit (even in the presence of long-range interactions), while 
it was argued that the oscillation amplitude may remain {\em finite} in three 
dimensions \cite{Lip}.
The authors of Ref.~\cite{Provata} considered a non-diffusive three-state 
model (each site can be empty, or occupied either by a predator or a prey) 
interacting according to a {\em cyclic} scheme. 
This system can in fact be mapped onto the so-called `rock-scissors-paper' (or 
three-state cyclic Lotka--Volterra) model \cite{Sigmund,cyclic}, well-known in
the field of game theory \cite{Sigmund}. 
The (mean-field) rate equations associated with that model actually admit a 
constant of motion \cite{Provata,Sigmund} which in turn implies cycles in the 
phase portrait, describing regular oscillations of the densities of predators 
and prey. 
However, numerical simulations of the stochastic version of that model on 
two-dimensional lattices led to a completely different behavior: 
The system was shown to display erratic oscillations whose amplitude vanished 
on a global scale for large lattices, but which were reported to persist on a 
smaller scale. 
These were explained in Ref.~\cite{Provata} as being associated with `small 
oscillators' (actually fluctuations) that are out of phase. 
Also, the fractal dimension of the patterns developed on the square lattice as
result of the spatial fluctuations of the reactants was investigated 
\cite{Provata}. We also would like to mention that Boccara et al. 
\cite{Boccara} studied a two-dimensional automaton network predator-prey model 
(a three-state system) with parallel updating for all the reactions except for 
`smart motion' (updated sequentially) of the predators (prey) which propagate 
toward the direction of highest (lowest) prey (predator) density. 
The authors of Ref.~\cite{Boccara} studied the phase portrait, finding a stable
coexistence state which may exhibit noisy cyclic behavior associated with 
complex patterns, and computing critical exponents, which in some regime are in
reasonable agreement with the DP values. 
Later, other authors \cite{Albano} considered the two-dimensional lattice-gas 
(with sequential updating) version of the model introduced by Boccara et al. 
and studied numerically its phase diagram and critical properties, finding 
again results consistent with the DP universality class. 
In addition, for the two-dimensional model of Refs.~\cite{Albano}, Rozenfeld 
and Albano argued that, in good agreement with mean-field results, there exists
a region of the phase diagram where the densities of species ``exhibit 
self-sustained oscillations'', with amplitudes that remain finite in the 
thermodynamic limit. 
As such a result was somewhat surprising, but could stem from the long-range 
interaction between predators and prey displayed in the model of 
Refs.~\cite{Albano}, Lipowski, willing also to test the general validity of the
scenario outlined in Refs.~\cite{Albano}, checked that the range of interaction
did {\em not} affect the characteristics of the oscillatory behavior displayed 
in two dimensions by the model of Ref.~\cite{Lip}: 
Actually, the amplitude of the oscillations was always found to vanish in the 
thermodynamic limit.

Before specifying further on two-species (stochastic) predator--prey models, 
for the sake of completeness we give a brief overview of some properties of 
the {\em multi-species} Lotka--Volterra rate equations.
In general, for $n$ particle or population species the latter read 
(with $i=1,\dots, n$) \cite{May1,Sigmund,Montroll}:
\begin{equation}
\label{genLV}
  \frac{d x_i(t)}{dt} = x_i(t) \biggl( r_i + \sum_{j=1}^{n} \alpha_{i,j} \,
  x_{j}(t) \biggr) ,
\end{equation}
where $x_i$ denotes the species $i$, the $r_i$ are the intrinsic growth 
($r_i > 0$) or decay ($r_i < 0$) rates, and $\alpha_{i,j}$ represents the 
{\em interaction} matrix that encodes the competition `reaction' between 
species $i$ and $j$.
For general $\alpha_{i,j}$ and an arbitrary number $n>2$ of species, many 
questions remain wide open.
One of the most intriguing (and less understood) features is the fact that the
deterministic equations (\ref{genLV}) may generate {\em chaotic} behavior 
already for only three species ($n=3$) \cite{Sigmund,May1}.
In the case where $\alpha_{i,j} = -\alpha_{j,i}$, and thus $\alpha_{i,i} = 0$ 
(which means that there is no nonlinear interaction within the same species),
it has been shown that Eq.~(\ref{genLV}) allows for a constant of motion 
(conserved first integral) \cite{Montroll}.
Nonetheless, also in this situation, for an even number of species with
$ n \geq 4$ it was demonstrated that Eq.~(\ref{genLV}) can display chaotic 
behavior resembling Hamiltonian chaos \cite{Picard}.
We also mention that when three or more species are in {\em cyclic} 
competition according to a dynamics described by Eqs.~(\ref{genLV}), i.e., the
rate equations are invariant under cyclic permutations of the species, quite
intriguing behavior may emerge: 
For some time it looks as if one species were bound to become the unique 
`survivor'; then its density drops and it is replaced with another apparently
dominant species, and after some time a third species seems to become 
dominant, and so on cyclically, involving all $n$ species 
\cite{Leonard,Sigmund}.

In the case of a food chain with $n$ components, where there is interaction 
(competition) among agents of the same species and where the first species 
serves as the prey for the second, which is the prey of the third and so on, 
the only nonzero entries of the interaction matrix of Eq.~(\ref{genLV}) are 
$\alpha_{i,i} < 0$, $\alpha_{i,i+1} < 0$, $\alpha_{i,i-1} > 0$ 
($\alpha_{1,0} = 0$), and also $r_1 > 0$, $r_i < 0$ for $i > 1$. 
In this case, it is known \cite{Sigmund} that the situation with $n = 2$ is 
generic and already captures the features of the multi-species system. 
In Sec.~II.C we shall discuss in detail the properties of a system 
[Eqs.~(\ref{EOMinter},\ref{EOMinter1})] which can be recast into the above 
two-species case.
Little is known as yet about spatial multi-species Lotka--Volterra systems 
defined on lattices, where spatial fluctuations generally invalidate (at least
in low dimensions) the predictions from the mean-field rate equations 
(\ref{genLV}). 
We note, however, that adding multiplicative noise as appropriate for the
existence of inactive, absorbing states transforms Eqs.~(\ref{genLV}) to the
Langevin equations for multi-species directed percolation processes, whose 
critical properties were shown by Janssen to be generically described by the 
DP universality class \cite{Janssen}.
A remarkable exception is the stochastic cyclic Lotka--Volterra model 
\cite{cyclic}, mimicking a simple cyclic food chain of length $n$, where the 
species $A_i$ ($i=1,\dots,n$) react according to the scheme 
$A_1 + A_2 \to 2 A_1$, 
$A_2 + A_3 \to 2 A_2, \dots, A_{n-1} + A_n \to 2 A_{n-1}$, 
$A_n + A_1 \to 2 A_n$. 
For this system, Frachebourg and Krapivsky \cite{cyclic} showed analytically 
(within a so-called Kirkwood decoupling scheme), and confirmed numerically, 
that in any dimension there is a critical number of species $n_c$ above which 
the system reaches a frozen steady state, i.e., there is {\em fixation} 
\cite{cyclic}, characterized by inert, non-fluctuating domains at whose 
interfaces all dynamics ceases (for example, in one dimension, 
$A_1 \dots A_1 A_3 \dots A_3 A_5 \dots A_5 A_2 \dots A_2 A_4 \dots A_4$).
In one dimension, the minimal number of species in order to have fixation is 
five; for $n < 5$ the systems coarsens: a large domain of a single species 
eventually spans the whole lattice.
In two and three dimensions it was found that $n_c = 14$ and $n_c = 23$,
respectively; for $n < n_c$, the steady state is reactive in $d = 2,3$.

All the above models are non-diffusive in the sense that there is no explicit 
mechanism allowing the mixing of the system: the agents are considered as 
{\em immobile} (but species may still spread because of the particle 
production processes, since new offspring have to be put to adjacent sites). 
Clearly, as noted by other authors \cite{Lip,Provata,Boccara,Albano,Droz}, a 
more realistic description of the predator--prey interaction should include 
the possibility for the agents to move. 
In fact, in ecosystems a prey tend to avoid the interaction with an incoming 
predator, while the predators aim to pursue the prey. 
In the absence of mixing processes, one can expect that the stochastic lattice
predator--prey model should display features like fixation (which is 
interesting but does not seem realistic from an ecological perspective).
One approach, followed here, is to allow the system to be mixed via particle 
(predators and prey) {\em diffusion}. 
Another approach, considered elsewhere for a model with next-nearest-neighbor 
interaction \cite{MGT1}, is to consider a nearest-neighbor exchange 
process (among any agents: predators, prey and empty sites) referred to as 
`stirring'. 
Interestingly, completely different results are obtained for systems mixed 
through diffusion or stirring, respectively: 
While in Sec.~III.C we shall explain that diffusion does not affect the 
critical and other generic properties of the system under consideration here, 
the exchange process is in fact capable of completely washing out the subtle 
correlations induced by long-range interactions. 
We discuss this latter issue in detail elsewhere \cite{MGT1}.

After this general discussion, let us anticipate the main results of this 
present work (of which a partial and brief account has recently been outlined in 
Ref.~\cite{MGT1}):
\begin{itemize}
\item We provide a comparison of various mean-field predictions with the
	results of numerical Monte Carlo simulations in dimensions 
	$1 \leq d \leq 4$, addressing the phase diagram, the structure of the 
	phase portrait, the existence and properties of the predator 
	extinction phase transition, and other issues (Secs.~II.C and III.C). 
\item We analytically derive some exact properties of the SLLVM (Sec.~III.B),
	quantitatively study the phase portrait and characterize the 
	properties of the intriguing spatial structures in the oscillatory
	regime of the active coexistence state by numerically computing 
	several correlation functions (Sec.~III.C).
\item We study the emergence of transient stochastic oscillations in the SLLVM
	and discuss the functional dependence of their characteristic 
	frequency as well as the dependence of their amplitude on the system
	size (Sec.~III.C). 
\item We provide a renormalization group argument, based on a field theory 
	representation of the corresponding master equation, that establishes
	that the active to absorbing extinction transition is indeed governed 
	by the directed percolation (DP) universality class (Sec.~III.D). 
\end{itemize}

Our results thus both confirm and supplement earlier work, where, as 
discussed above, some of these issues have already been considered for 
related, but significantly different, systems such as the four-state models of 
Refs.~\cite{Lip,TiborDroz}, or the cyclic three-state model of 
Ref.~\cite{Provata}.
In particular, we shall show that the mean-field rate equations 
[Eqs.~(\ref{EOMinter},\ref{EOMinter1})] for the model under consideration here
already provide a {\em qualitatively} (albeit {\em not} quantitatively) 
correct description of the behavior of the corresponding stochastic lattice 
system, of which they therefore capture the essential features (in dimensions 
$d > 1$). 

The organization of this article is the following: 
Section II is devoted to a review of the properties of the deterministic 
two-species LVM. 
Section II.A covers the basic properties of the original Lotka--Volterra
coupled rate equations (see Ref.~\cite{Murray}, Vol.~I, Chap.~3). 
The fundamental features of the corresponding zero-dimensional stochastic 
model are reviewed in Sec.~II.B \cite{Haken,McKane}.
In Secs.~II.C and II.D, we consider the LVM rate equations subject to finite 
carrying capacities, and its spatial extension with diffusive particle 
propagation \cite{Dunbar} (see also Ref.~\cite{Murray}, Vol.~II, Chap.1).
Section III is devoted to the two-species stochastic lattice Lotka--Volterra 
model (SLLVM), as introduced in Sec.~III.A. 
Some exact properties of the SLLVM are discussed in Sec.~III.B. 
Section III.C is devoted to the results from our Monte Carlo simulations of 
the SLLVM: 
Dynamical features in the active phase as well as the critical properties near
the predator extinction threshold are presented and discussed in detail here. 
In Sec.~III.D, we present a field-theoretic analysis of the critical 
properties of the SLLVM. 
Section IV is devoted to our conclusions.

\section{Preliminaries: generic properties of the Lotka--Volterra model (LVM)
	and mean-field theory}

\subsection{The two-species Lotka--Volterra rate equations}

Following Lotka and Volterra's original work \cite{Lotka,Volterra}, we 
consider two chemical or biological species, the `predators' $A$ and `prey' 
$B$, in competition: the predators consume the prey and {\em simultaneously}
reproduce with rate $\lambda > 0$. 
In addition, the prey may reproduce with rate $\sigma$ and the predators are 
assumed to spontaneously die with rate $\mu$. 
Neglecting any spatial variations of the concentrations, which we shall denote
by $a({\bm x},t)$ and $b({\bm x},t)$ for species $A$ and $B$, respectively, 
the heuristic mean-field rate equations for this reaction model are given by 
the classical coupled nonlinear Lotka--Volterra (LV) differential equations 
\cite{Lotka,Volterra}:   
\begin{eqnarray}
\label{clotvol}
  && \dot{a}(t) = \lambda \, a(t) \, b(t) - \mu \, a(t) \ , \\ 
\label{blotvol}
  && \dot{b}(t) = \sigma \, b(t) - \lambda \, a(t) \, b(t) \ ,
\end{eqnarray}
where the dot denotes the time derivative.
Note that within this mean-field approximation, we may view the parameters 
$- \mu = \sigma_A - \mu_A$ and $\sigma = \sigma_B - \mu_B$ as the net 
population growth rates for competing birth/death processes ($A \to A + A$ 
and $A \to \oslash$, where $\oslash$ denotes an empty `spot') with rates 
$\sigma_A$ and $\mu_A$, respectively, and similarly for species $B$.
For $\mu > 0$ and $\sigma < 0$, clearly both populations will die out
exponentially, whereas $\mu < 0$ and $\sigma > 0$ leads to unbounded
population growth.
Therefore, interesting feedback interactions between the `prey' $B$ and the
`predators' $A$, which would become extinct in the absence of the prey, 
occur only if both $\mu$ and $\sigma$ (as well as $\lambda$) are positive.

The coupled deterministic evolution equations (\ref{clotvol}), (\ref{blotvol})
have as stationary states (fixed points) $(a^*,b^*) = (0,0)$ (extinction), 
$(0,\infty)$ (predators extinct, Malthusian prey proliferation), and 
$(a_c,b_c) = (\sigma/\lambda , \mu/\lambda)$ (species coexistence).
For positive $\mu$ and $\sigma$, the `trivial' steady states with $a = 0$ and 
$b = 0$ and $\infty$ are both linearly unstable [in the absence of predation,
$\lambda = 0$, $(0,\infty)$ is stable].

Notice, however, that they both constitute {\em absorbing} stationary states, 
since neither can be left through the involved reactions alone.
Linearizing about the nontrivial coexistence stationary state, 
$\delta a(t) = a(t)-a_c$, $\delta b(t) = b(t)-b_c$, one obtains to first order
in $\delta A$ and $\delta B$: $\delta \dot{a}(t) = \sigma \, \delta b(t)$, and
$\delta \dot{b}(t) = - \mu \, \delta a(t)$. 
The eigenvalues of the corresponding Jacobian (also occasionally termed 
stability or community matrix) are $\pm i \sqrt{\mu \sigma}$, which suggests 
purely oscillatory kinetics in the vicinity of the neutrally stable fixed 
point (center singularity) $(a_c,b_c)$.
Indeed, one finds the general periodic solutions $\delta a(t) = 
\delta a(0) \, \cos\left( \sqrt{\mu \sigma} \, t \right) + \delta b(0) \, 
\sqrt{\sigma/\mu} \, \sin\left( \sqrt{\mu \sigma} \, t \right)$, and
$\delta b(t) = - \delta a(0) \, \sqrt{\mu/\sigma} \, 
\sin\left( \sqrt{\mu \sigma} \, t \right) + \delta b(0) \, 
\cos\left( \sqrt{\mu \sigma} \, t \right)$, with characteristic frequency 
$\omega = \sqrt{\mu \sigma}$.

Going beyond linear stability analysis, one easily confirms that the quantity 
\begin{equation}
\label{clvcons}
  K(t) = \lambda [a(t) + b(t)] - \sigma \ln{a(t)} - \mu \ln{b(t)}  
\end{equation}
represents a conserved first integral for any phase space trajectory,
$\dot{K}(t) = 0$.
Quite generally, therefore, Eqs.~(\ref{clotvol}), (\ref{blotvol}) yield 
periodic oscillations of both species concentrations, whose amplitudes (and in
the nonlinear regime, also whose frequencies) are determined by the 
{\em initial} conditions $a(0)$ and $b(0)$, and according to 
Eq.~(\ref{clvcons}) neither $a(t)$ nor $b(t)$ can ever vanish.
The emergence of purely oscillatory kinetics in this classical Lotka--Volterra
model, irrespective of the involved reaction rates, and in character only 
determined by the initial conditions, is clearly not a realistic feature 
\cite{Neal,Murray}.

\subsection{The zero-dimensional stochastic Lotka--Volterra model}

Instead of the regular cycles, completely determined by the {\em initial}
conditions, predicted by the rate equations (\ref{clotvol}), (\ref{blotvol}), 
one would more realistically typically expect stable stationary states with 
fixed concentrations, and/or the possibility of extinction thresholds.
Indeed, the conservation law for $K(t)$ and the related property that the
eigenvalues of the linearized kinetics near coexistence are purely 
imaginary, constitute very special features of the {\em deterministic} model
equations --- there is in fact no underlying physical background for the 
conserved quantity (\ref{clvcons}).
Correspondingly, the above center singularity is {\em unstable} with respect
to perturbations: namely, either with respect to introducing modifications of 
the model equations, spatial degrees of freedom, and/or stochasticity.
For obviously, when the number of predators becomes very low, a chance 
fluctuation may lead the system into the absorbing state with $a = 0$.
Consider the zero-dimensional stochastic Lotka--Volterra model that is
governed by the following master equation, stating the gain and loss balance 
for the temporal evolution of the probability of finding $A$ predators and 
$B$ prey in the system, 
\begin{eqnarray}
\label{0dstolv}
  &&\dot{P}(A,B;t) = \lambda (A-1) \, (B+1) \, P(A-1,B+1;t) \nonumber \\
  &&\ + \mu \, (A+1) \, P(A+1,B;t) + \sigma \, (B-1) \, P(A,B-1;t) \nonumber\\
  &&\ - (\mu \, A + \sigma \, B + \lambda \, A \, B) \, P(A,B;t) \ .
\end{eqnarray}
In this description, the system has {\em discrete} degrees of freedom, and it 
can be verified that its only stable stationary state ($\dot{P} = 0$) is 
$P_s(A=0,B=0) = 1$ and $P_s(A\not=0,B\not=0) = 0$ \cite{Haken}.
Therefore, {\em asymptotically} as $t \to \infty$ the empty state will be 
reached, which is {\em absorbing}, since all processes cease there, and no
fluctuation can drive the system out of it anymore.
However, at {\em finite} times such stochastic Lotka--Volterra systems still 
display quite intriguing dynamics: namely, inevitable fluctuations tend to 
push the system away from the trivial steady state and induce erratic 
population oscillations that almost resemble the deterministic cycles.
This `resonant amplification mechanism' is always present in finite 
populations and can significantly delay extinction \cite{McKane}.
We shall later, in Sec.~III.C, discuss the analog of this mechanism in the
spatially extended models.

\subsection{Mean-field rate equations with finite carrying capacities}

In the ecological and biological literature, at the rate equation level, 
population models such as (\ref{clotvol}), (\ref{blotvol}) are rendered 
more `realistic' by introducing {\em growth-limiting} terms that describe 
a finite `carrying capacity' \cite{Neal,Murray}.
In a similar manner, in a spatial system one may need to take into account 
that the {\em local} population densities cannot exceed some given, bound 
value, which typically depends on external factors; this amounts to 
introducing {\em spatial constraints}, e.g., in a lattice model, restrictions
on the maximum possible occupation number on each site.
There are various possibilities to introduce carrying capacities; here we 
consider the very natural choice of limiting the effective reproduction term 
in Eq.(\ref{blotvol}) for the prey, in the form 
$\sigma \, b(t) \left[ 1 - \zeta^{-1} \, a(t) - \rho^{-1} \, b(t) \right]$, 
where $0 \leq \zeta^{-1} \leq \rho^{-1} \leq 1$.
In the absence of the predators, $\rho$ represents the prey carrying capacity.
In the presence of predators, it is further diminished by the cross-species 
interactions. 
In the lattice model, these choices reflect the fact that prey reproduce only
when an `empty spot' is available in its immediate vicinity.
The resulting rate equations now read (with $0 < \rho \leq \zeta$):
\begin{eqnarray}
\label{EOMinter}
  \dot{a}(t) &=& a(t) \left[ \lambda \, b(t) - \mu \right] , \\
\label{EOMinter1}
  \dot{b}(t) &=& \sigma b(t) \left[ 1 - \zeta^{-1} \, a(t) - \rho^{-1} \, b(t)
  \right] - \lambda a(t) \, b(t) \ . \quad
\end{eqnarray}
As shown in Sec.~III.B, when $\zeta = \rho = 1$,  Eqs.~(\ref{EOMinter}) and
(\ref{EOMinter1}) can be interpreted as the mean-field versions of the
{\em exact microscopic} equations derived from a stochastic lattice 
formulation (based on the corresponding master equation) where each lattice
site may at most be occupied by a single particle. 
Obviously, in this case $0 \leq a(t) + b(t) \leq 1$ (provided that this
inequality holds initially at time $t = 0$).
[Indeed, even when Eq.~(\ref{EOMinter1}) cannot be related to some 
microscopic dynamics, it is readily verified that still 
$0 \leq \zeta^{-1} \, a(t) + \rho^{-1} \, b(t) \leq 1$, provided one starts 
with a `physical' initial condition, i.e., 
$0 \leq \zeta^{-1} \, a(0) + \rho^{-1} \, b(0) \leq 1$ and, in addition, 
$0 < \rho \leq \zeta$ holds.]

The coupled rate equations (\ref{EOMinter}) and (\ref{EOMinter1}) have three 
fixed points. 
The two obvious ones are $(a^*_1,b^*_1) = (0,0)$ (total population extinction)
and $(a^*_2,b^*_2) = (0,\rho)$, corresponding to a system filled with prey up
to its carrying capacity.  
The only nontrivial fixed point, associated with the coexistence of both 
populations, is $(a^*_3,b^*_3)$, with
\begin{eqnarray}
\label{NTinter}
  a^*_3 = \frac{\zeta \, \sigma}{\zeta \, \lambda + \sigma} 
  \left( 1 - \frac{\mu}{\lambda \, \rho} \right) \ , \quad 
  b^*_3 = \frac{\mu}{\lambda} \ .
\end{eqnarray}
Species coexistence is obviously possible only when $\lambda > \mu / \rho$.
For fixed predator death rate $\mu$ and prey carrying capacity $\rho$, the
predator population dies out if $\lambda \leq \lambda_c = \mu / \rho$, which
represents the predator {\em extinction threshold}.
For $\lambda \to \lambda_c$ from above, the stationary predator density 
tends to zero continuously; hence predator extinction constitutes a continuous
nonequilibrium phase transition from the active coexistence phase to an
inactive, absorbing state: once all predators are gone, no mechanism, not even
stochastic fluctuations, allows them to ever reappear in the system.  

Before proceeding with the linear stability analysis of Eqs.~(\ref{EOMinter})
and (\ref{EOMinter1}), we remark that there exists a {\em Lyapunov function} 
$V(a,b)$ \cite{Sigmund,Murray} associated with those equations.
In fact, with 
\begin{eqnarray}
\label{Lyap}
  V(a,b) &=& \lambda \left[ b^*_3 \ln{b(t)} - b(t) \right] \nonumber \\ 
  &&+ \left( \lambda + \sigma / \zeta \right) \left[ a^*_3 \ln{a(t)} - a(t) 
  \right] \ ,
\end{eqnarray}
we have $\dot{V}(a,b) = 
\frac{\lambda \sigma}{\rho} \left[ b^*_3 - b(t) \right]^2 \geq 0$ and 
$\dot{V}(a^*_3,b^*_3) = 0$.
According to Lyapunov's theorem, every flow $(a(t),b(t))$ is contained in 
$\{ (a,b)| \dot V(a,b) = 0 \}$. 
Therefore, since $V(a,b) > 0 \quad \forall (a,b) \neq (a^*_3,b^*_3)$ and the 
neighborhood of $(a^*_3,b^*_3)$ represents an invariant subset  (the so-called 
$\omega$-limit property in Chap.~2 of Ref.~\cite{Sigmund}), it follows that
$(a^*_3,b^*_3)$ is indeed {\em globally stable} (when physically accessible, 
i.e., for $\lambda >\mu/\rho$). 
Only when $\rho = \infty$, as is the case in the classical LV equations, 
$V(a,b) = 0$, and in this situation $(a^*_3, b^*_3) \to \left( 
\frac{\zeta \, \sigma}{\zeta \, \lambda + \sigma} , \frac{\mu}{\lambda} 
\right)$ is {\em not} globally stable.

There is general no methods to find a Lyapunov function (provided it even
exists) associated with a given set of coupled ordinary differential 
equations. One thus often relies on generic mathematical results, such as 
Kolmogorov's theorem (see Refs.~\cite{Kolmogorov,May1} and references therein) 
and the so-called Bendixson--Dulac test (Refs.~\cite{nonlin,Sigmund}) to 
establish the existence of a stable fixed point or limit cycle. 
As explained in Appendix A, Kolmogorov's theorem does not apply to 
Eqs.~(\ref{EOMinter}), (\ref{EOMinter1}), while the Bendixson--Dulac test
yields that these equations do {\em not} admit periodic orbits (as long as 
there is a finite carrying capacity, i.e. $\rho < \infty$).

We now proceed with an analysis of the properties of the various fixed points 
of the rate equations (\ref{EOMinter}) and (\ref{EOMinter1}) by means of 
linear stability analysis \cite{Murray}.
To this end, we need to diagonalize the Jacobian matrix ($i=1,2,3$)
\begin{eqnarray}
\label{Jacinter}
  J= \left( \begin{array}{cc} \lambda \, b^*_i - \mu & \lambda \, a^*_i \\
 - \left( \frac{\sigma}{\zeta} + \lambda \right) b^*_i & \sigma \left( 1 -
 \frac{2 b^*_i}{\rho} \right) - \left( \frac{\sigma}{\zeta} + \lambda \right)
 a^*_i \\ \end{array} \right) . \quad
\end{eqnarray}
This gives the eigenvalues associated with the fixed point $(0,0)$ to be 
$\epsilon_+ = \sigma$ and $\epsilon_- = - \mu$, which implies that the empty 
lattice fixed point is a saddle point (unstable in the $b$ direction), for 
any value of $\zeta$ and $\rho$.
For the fixed point $(0,\rho)$ (system filled with prey), the eigenvalues 
read $\epsilon_+ = \rho \, \lambda -\mu$ and $\epsilon_- = -\sigma$. 
This means that $(0,\rho)$ is a {\em stable node} provided 
$\lambda < \lambda_c = \mu / \rho$. 
When $\lambda > \lambda_c$, where the nontrivial fixed point (\ref{NTinter})
yields a positive predator density, $\epsilon_+$ becomes positive, and 
$(0,\rho)$ turns into a saddle point (unstable in the $a$ direction). 
This result also confirms that in the absence of any local density 
restriction on the prey, i.e., in the limit $\rho \to \infty$, the fixed 
point $(0,\rho) \to (0,\infty)$ becomes unstable for any $\lambda > 0$.

For the nontrivial fixed point given by Eq.~(\ref{NTinter}), the corresponding
eigenvalues are
\begin{equation}
\label{eigNTint}
  \epsilon_\pm = - \frac{\sigma \, \mu}{2 \lambda \, \rho} \left[ 1 \pm
  \sqrt{1 - \frac{4 \lambda \, \rho}{\sigma} \left( 
  \frac{\lambda \, \rho}{\mu} - 1 \right)} \right] \ ,
\end{equation}
and the different emerging scenarios can be summarized as follows:
\begin{itemize}
\item for $\lambda \in \left] \lambda_c , \lambda_s \right]$ or 
      $\sigma > \sigma_s$, where $\lambda _c = \mu / \rho$ and
      $\lambda_s = \frac{\mu}{2 \rho} \left( 1 + \sqrt{1 + \frac{\sigma}{\mu}}
      \right)$ and $\sigma_s = 4 \lambda \, \rho \left(
      \frac{\lambda \, \rho}{\mu} - 1 \right) > 0$, the eigenvalues are real 
      with $\epsilon_\pm < 0$: the fixed point is a {\em stable node};
\item for $\lambda \in \left] \lambda_s , \infty \right[$ or 
      $\sigma < \sigma_s$, the eigenvalues $\epsilon_\pm$ have both real and 
      imaginary parts; in this case $\Re (\epsilon_\pm) < 0$ and 
      $|\Im (\epsilon_\pm)| \neq 0$, provided that $\rho < \infty$
      ($\mu \, \sigma / \rho \, \lambda > 0$): the nontrivial fixed point 
      (\ref{NTinter}) is an {\em attractive focus};
\item if $\rho \to \infty$ and/or if $\mu \, \sigma / \rho \, \lambda \to 0$ 
      (with finite $\mu$ and $\sigma$), the real part of the eigenvalues
      vanishes, whence $\epsilon_\pm \to \pm i \, \sqrt{\mu \, \sigma}$. 
      In this extreme situation the nontrivial fixed point at $\rho = \infty$ 
      becomes a center singularity, and we encounter periodic cycles in the 
      phase portrait. 
      In the case when $\rho$ is finite and 
      $\mu \, \sigma / \rho \, \lambda \to 0$, the fixed point evolves 
      towards one of the phase space boundaries, and no cyclic behavior can be 
      established.
\end{itemize}
It is worthwhile noticing that these scenarios quantitatively differ from 
those predicted by the rate equations in the models of 
Refs.~\cite{Lip,Provata} and turn out to be essentially independent of 
the actual value of $\zeta$, which is the parameter that controls the spatial 
restrictions of the predators on the prey population.

As a result of this discussion, on the rate equation level, it turns out 
that the growth-limiting constraints (as long as $\rho^{-1} > 0$) generically 
invalidate the classical Lotka--Volterra picture. 
Interestingly, the density restrictions induce a {\em continuous} active to
absorbing phase transition, namely an extinction threshold for the predator 
population at $\lambda_c = \mu / \rho$, which can be accessed by varying the
reaction rates. At mean-field level (then confirmed by numerical simulations 
of the related stochastic models), extinction phase transitions were also
reported in Refs.~\cite{Tome,TiborDroz,Albano}. 
In the vicinity of the phase transition at $\lambda_c$ (with $\sigma$ and 
$\mu$ held fixed), the density of predators approaches its stationary value 
linearly according to Eqs.~(\ref{EOMinter}), (\ref{EOMinter1}): 
$a^*_3 \sim (\lambda -\lambda_c)^{\beta_{\rm MF}}$, with $\beta_{\rm MF} = 1$.
Moreover, depending on these rates, the only stable fixed point, 
corresponding to the coexistence of both populations of predators and prey, 
is either a node or a focus, and therefore approached either directly or in 
an oscillatory manner.
Near the predator extinction threshold, the active fixed point is a node;
deeper in the population coexistence phase, it changes its character to a
focus. 
In Sec.~III.C, we shall test the validity of the results arising from the 
deterministic mean-field rate equations (\ref{EOMinter}) and (\ref{EOMinter1})
by considering stochastic lattice predator--prey models defined on lattices 
and formulated in a  microscopic setting (i.e., starting from stochastic 
dynamical rules) taking into account internal noise (fluctuations).

\subsection{The deterministic reaction--diffusion equations (with finite 
	carrying capacities)}

To account for the spatial structure on a rate equation level, and 
guaranteeing asymptotic stability of the coexistence state with both 
nonvanishing predator and prey populations, one may introduce spatial degrees 
of freedom.
This effectively allows the prey to `escape' via diffusion, which in turn 
requires the predators to `pursue' them, thus effectively generating delay 
terms in the kinetics that stabilize the nontrivial steady state 
\cite{Murray}.
As previously, this can be done in a heuristic fashion with 
finite carrying capacity $\rho$ and additional growth-limiting term $\zeta$ 
for the prey, and adding diffusive terms $\nabla^2 a({\bm x},t)$ and 
$\nabla^2 b({\bm x},t)$ to Eqs.~(\ref{EOMinter}) and (\ref{EOMinter1}), with 
diffusivities $D_{A}$ and $D_B$:
\begin{eqnarray}
\label{dlotvol}
  \frac{\partial a({\bm x},t)}{\partial t} &=& D_A \, \nabla^2 a({\bm x},t)
  + \lambda \, a({\bm x},t) \, b({\bm x},t) - \mu \, a({\bm x},t) \ , 
  \nonumber \\ 
  \frac{\partial b({\bm x},t)}{\partial t} &=& D_B \, \nabla^2 b({\bm x},t)
  - \lambda \, a({\bm x},t) \, b({\bm x},t) \\ 
  &&+ \sigma \, b({\bm x},t) \left[ 1 - \zeta^{-1} \, a({\bm x},t)
  - \rho^{-1} \, b({\bm x},t) \right] \ . \nonumber
\end{eqnarray}
For instance, it is straightforward to construct one-dimensional wavefront 
solutions to the deterministic coupled reaction--diffusion equations 
(\ref{dlotvol}) of the form $a(x,t) = a(x+ct)$ and $b(x,t) = b(x+ct)$, which 
interpolate between the stationary states $(0,\rho)$ and $(a^*_3,b^*_3)$.
In fact, depending on the rate parameters, there exist two types of such
travelling waves of `pursuit and evasion', namely either with monotonic or
oscillatory approach to the stable state \cite{Dunbar}; these correspond to 
the different scenarios discussed in the previous Sec.~II.C.
For $D_B = 0$, one finds a minimum wavefront propagation velocity
$c \geq \left[ 4 D_A \left( \lambda \, \rho - \mu \right) \right]^{1/2}$ 
(see Ref.~\cite{Murray}, Vol.~II, Chap.~1.2.).

As we shall explain in Sec.~III.C (see Fig.~\ref{2d_snap}), such `pursuit and 
evasion' waves in the predator and prey density fields arise naturally in the 
SLLVM in a certain region of parameter space, even in the absence of explicit 
diffusion of either species.  
We also mention that the problem of velocity selection for reaction fronts 
starting from a microscopic description, e.g., from the associated master 
equation, for the underlying stochastic processes is a rather subtle issue 
\cite{Riordan,Levine}.
For some two-state models, a field-theoretic representation has been useful 
to derive a stochastic differential equation that properly represents the
underlying stochastic process \cite{Levine}. 
However, to the best of our knowledge no similar treatment has as yet been
devised for Lotka--Volterra type interactions.

\section{The Stochastic Lattice Lotka--Volterra Model (SLLVM)}

In the remainder of this paper, we study and carefully discuss the role of
spatial structure and intrinsic stochastic noise on the physical properties of
systems with Lotka--Volterra type predator--prey interaction, starting from
a microscopic (stochastic) formulation. 
We shall compare the results of the SLLVM with those predicted by the rate 
equations (\ref{EOMinter}), (\ref{EOMinter1}) and will show that there is 
qualitative agreement for many overall features (in dimensions $d > 1$), but
there are also important differences.
We shall investigate the site-restricted stochastic version of the lattice 
Lotka--Volterra model.
Starting from the master equation governing its stochastic kinetics, we shall
employ numerical Monte Carlo simulations as well as field-theoretic arguments.

\subsection{The SLLVM (with site restrictions) as a {\em stochastic} 
	reaction--diffusion model}
	 
We define the SLLVM with site restrictions as a {\em microscopic} 
reaction--diffusion system on a periodic hypercubic lattice of linear size 
$L$, whose sites ${\bm j}$ are labeled by their components 
${\bm j} = (j_1, \dots, j_d)$, where $d$ denotes the spatial dimension, and 
the unit vectors are represented by ${\bm e}_i$ for $i \in (1,\dots,d)$. 
Each lattice site can either be empty ($\oslash$), occupied by a `predator' 
($A$ particle) or by a `prey' ($B$ particle). 
Multiple occupancy is {\em not} allowed, and the stochastic rules determining 
the system's dynamics are defined as follows:
\begin{itemize}
\item $A \xrightarrow[\mu]{}\oslash$: death of a predator with rate $\mu$;
\item $A \oslash \xrightarrow[D/z]{}\oslash A$ and $B \oslash 
	\xrightarrow[D/z]{} \oslash B$: nearest-neighbor hopping (diffusion) 
        with rate $D / z$;
\item $B \oslash \xrightarrow[\sigma/z]{} B B$: branching (offspring 
	generation) of a prey with rate $\sigma / z$;
\item $A B \xrightarrow[\lambda/z]{} A A$: predation interaction: a predator 
        consumes a prey and produces an offspring with rate $\lambda / z$.
\end{itemize}
In the above rules the quantity $z = 2 d$ represents the lattice
coordination number; all processes occur isotropically in space, i.e., 
there is no spatial bias in the reaction rates. 
We also notice that the process with rate $\mu$ represents a single-site 
reaction, whereas the processes with rates $D$, $\sigma$, and $\lambda$ 
describe nearest-neighbor two-site reactions.

Microscopically, each configuration ${\cal C}={\cal C}_{\{A,B,\oslash\}}$ of 
the system at time $t$ is characterized by a probabilistic weight 
$P({\cal C},t)$. 
The temporal evolution of this probability distribution is governed by a 
master equation:
\begin{eqnarray}
\label{master}
  \dot{P} ({\cal C},t) &=& \sum_{{\cal C}'\neq {\cal C}} 
  W({\cal C}' \to {\cal C}) \, P({\cal C}',t) \nonumber \\
  &&- \sum_{{\cal C}'\neq {\cal C}} W({\cal C}\to {\cal C}') \, P({\cal C},t)
  \ ,
\end{eqnarray}
where the transition from the configuration ${\cal C}'$ to ${\cal C}$ (during 
an infinitesimal time interval $dt$) occurs through a single reaction event 
with nonzero rate $W({\cal C}' \to {\cal C})$.
The first term on the right-hand-side of Eq.~(\ref{master}) is the `gain term'
accounting for contributions entering the configuration ${\cal C}$, while the 
second (`loss') term captures the processes leaving ${\cal C}$. 
Of course, the configurations ${\cal C}$ and ${\cal C}'$, as well as the 
transition rates, should be compatible with the processes underlying the 
dynamics. 
For instance, in one dimension, the configuration ${\cal C} = \{ 
B,\,A,\,\oslash,\,B,\, A,\,\oslash,\,\dots,\,A, \,\oslash,\,A,\,\oslash,\,B 
\}$ is compatible with ${\cal C}'= \{A,\,A,\,\oslash,\,B,\,A,\,\oslash,\,
\dots,A,\,\oslash,\,A,\,\oslash,\,B \}$, and in this case the transition 
${\cal C} \to {\cal C}'$ occurs with a rate 
$W({\cal C} \to {\cal C}') = \lambda / z$. 
Specifically, to account for the site restriction and the fact that we are 
dealing with a three-state model, the master equation can be rewritten in a 
matrix form by introducing suitable $3 \times 3$ operators, which are the 
direct generalization of Pauli's spin-$1/2$ operators (see, e.g., 
Ref.~\cite{3S}).
Within this spin-like reformulation, which is by now standard in the study of 
reaction--diffusion systems (see, e.g., Ref.~\cite{RD,DDS} for reviews), the 
master equation (\ref{master}) can formally be rewritten as an `imaginary-time
Schr\"odinger' equation where the `stochastic Hamiltonian' $H$, which is the 
Markovian generator, is in general not Hermitian. 
Taking advantage of such a reformulation, the equations of motion of all the 
observables, e.g., the density of particles, correlation functions, etc. can 
be obtained in a systematic algebraic fashion using the corresponding quantum 
physical Heisenberg picture (see, e.g., Ref.~\cite{RD}). 
In this language, the equation of motion of the average value of an observable
of interest, say  ${\cal O}$ (density, correlator, \dots), reads 
$\frac{d}{dt} \langle {\cal O}(t) \rangle = \sum_{\cal C} {\cal O}(t) \,
P({\cal C}, t) = \langle [H,{\hat {\cal O}}(t)] \rangle$, where the square 
bracket denotes the usual commutator and ${\hat {\cal O}}$ is the operator 
whose eigenvalue is ${\cal O}$. 

In addition to allowing us to derive exact properties of the phase portrait of 
the SLLVM (see below), the stochastic Hamiltonian reformulation of the master 
equation is the most suitable approach on which to build a field-theoretic
analysis of the critical properties of the system. Such a treatment is the 
scope of Section  III.D below.

\subsection{SLLVM equations of motion and some exact properties}

Let us now formulate the stochastic equation of motion for the density of the 
$A$ and $B$ particles, denoted respectively as before 
$a({\bm j},t) = \langle n_{\bm j}^A(t) \rangle$ and 
$b({\bm j},t) = \langle n_{\bm j}^B(t) \rangle$. 
The stochastic variable $n_{\bm j}^A$ ($n_{\bm j}^B$) represents the 
occupation number at site ${\bm j}$ by $A$ $(B)$ particles: 
$n_{\bm j}^A = 1 \; (n_{\bm j}^B = 1)$ if the site ${\bm j}$ is occupied by a 
predator (prey), and $0$ otherwise. 
Obviously, it follows that $\langle n_{\bm j}^{\oslash}(t) \rangle = 
1 - \langle n_{\bm j}^{A}(t) \rangle - \langle n_{\bm j}^{B}(t) \rangle$. 
Considering a translationally invariant system, it is then straightforward to
obtain the following {\em exact} equations of motion for the concentrations of
the predators and the prey from the master equation (\ref{master}):
\begin{eqnarray}
\label{EOMAhom}
  \dot{a}(t) &=& \lambda \, c_{AB}(t) - \mu \, a(t) \ , \\
\label{EOMBhom}
  \dot{b}(t) &=& \sigma \left[ b(t) - c_{BB}(t) - c_{AB}(t) \right] 
  - \lambda \, c_{AB}(t) \ ,
\end{eqnarray}
where $c_{AA}(t) = \langle n_{\bm j}^{A} n_{\bm{j+e_i}}^{A} \rangle(t)$,
$c_{BB}(t) = \langle n_{\bm j}^{B} n_{\bm{j+e_i}}^{B} \rangle(t)$, and 
$c_{AB}(t) = \langle n_{\bm j}^{A} n_{\bm{j+e_i}}^{B} \rangle(t) = 
\langle n_{\bm j}^{B} n_{\bm{j+e_i}}^{A}\rangle(t)$ represent the two-point 
correlation functions. 
Notice that the diffusion rate $D$ and the coordination number $z$ do not 
appear explicitly in Eqs.~(\ref{EOMAhom}) and (\ref{EOMBhom}). 
However, they would enter the equations of motion for the two-site probability
distributions, i.e., the correlators $c_{AB}(t)$ and $c_{BB}(t)$.

It is clear from these equations of motion that the quantity $K$ in 
Eq.~(\ref{clvcons}) is no longer a first integral of the motion of the 
stochastic model (with the site restrictions, this is even true on the 
mean-field level, as we saw in Sec.~II.C). 
Even though it is not possible to solve Eqs.~(\ref{EOMAhom}) and 
(\ref{EOMBhom}) in a closed form, owing to the emerging infinite hierarchy of 
higher-order correlations, we can still obtain some useful and nontrivial 
information on the phase portrait. 
Let us denote by $a^*$ and $b^*$ the stationary concentrations of the 
predators and the prey, respectively, and by $c_{BB}^*$ and $c_{AB}^*$ the 
stationary values of the correlators from Eqs.~(\ref{EOMAhom}) and 
(\ref{EOMBhom}). 
As the site occupation number restrictions imply 
$0 \leq a(t), b(t), c_{BB}(t), c_{AB}(t) \leq 1$, we thus have 
$0 \leq \mu a^* = \lambda c_{AB}^*$ and $0 \leq b^* - 
\frac{\sigma + \lambda}{\sigma \, \lambda} \, \mu \, a^* = c_{BB}^* \leq 1$. 
Thus, as a direct consequence of our reformulation of the problem, we arrive 
at the following inequalities, which considerably restrict the physically 
available phase portrait:
\begin{eqnarray}
\label{simplex}
  0 & \leq & a^* \leq {\rm min} \, \left(\frac{\lambda}{\mu}, 1 \right) ; 
  \quad 0 \leq b(t) \leq 1 \ , \nonumber \\
  0 & \leq & a(t) + b(t) \leq 1 \ , \nonumber \\
  0 & \leq & b^* - \frac{\sigma + \lambda}{\sigma \, \lambda} \, \mu \, a^* 
  \leq 1 \ .
\end{eqnarray}
We emphasize that the inequalities (\ref{simplex}) are {\em exact} and 
obtained from very general considerations starting from the master equation. 
In this sense they intrinsically account for the {\em spatial} and 
{\em stochastic} nature of the underlying reaction--diffusion system.
Upon ignoring any spatial fluctuations and correlations, which amounts to
assuming the factorizations $c_{AB}(t) = a(t) \, b(t)$ and 
$c_{BB}(t) = b(t)^2$, after substitution into Eqs.~(\ref{EOMAhom}) and 
(\ref{EOMBhom}) one recovers the deterministic mean-field rate equations 
(\ref{EOMinter}) and (\ref{EOMinter1}) with $\rho = \zeta = 1$. 
This implies that the site restrictions on a mean-field level correspond to a 
finite prey carrying capacity.

\begin{figure}[!t]
\includegraphics[angle=-90,width=8.6cm]{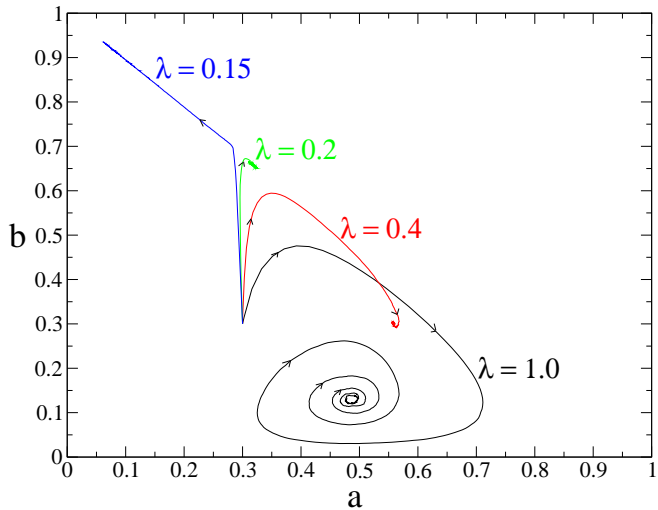}
\caption{(Color online.) Typical trajectories above in the predator/prey 
coexistence phase depicting the phase portrait for the NN model on a 
$(512 \times 512)$ lattice.
All runs start from random initial configuration with $a(0)=b(0)=0.3$ and fixed
rates $D=0$, $\sigma=4.0$, $\mu=0.1$, and $\lambda=0.15, 0.20, 0.40, 1.0$, 
respectively. For high values of $\lambda$ we observe the typical spirals (the 
fixed point is a focus) in phase space, while for small values of $\lambda$ 
(typically $\lambda < 0.4$) the fixed point is a node.}
\label{2d_traj}
\end{figure}

\begin{figure*}
\includegraphics[width=17.9cm]{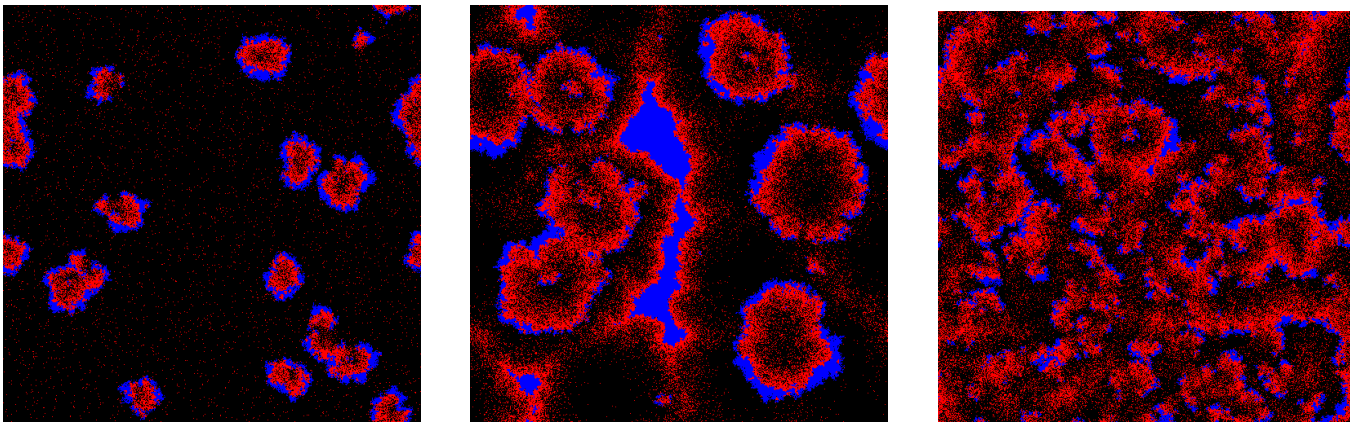}
\caption{(Color online.) Snapshots of the time evolution (time increases from 
left to right) of the two-dimensional SLLVM model in the species coexistence 
phase, when the fixed point is a focus. The red, blue, and dark dots 
respectively represent the prey, predators, and empty lattice sites. 
The rates here are $D = 0$, $\sigma = 4.0$, $\mu = 0.1$, and $\lambda = 2.2$. 
The system is initially homogeneous with densities $a(0) = b(0) = 1/3$ and the 
lattice size is $512 \times 512$.}
\label{2d_snap}
\end{figure*}

\begin{figure*}
\includegraphics[width=17.9cm]{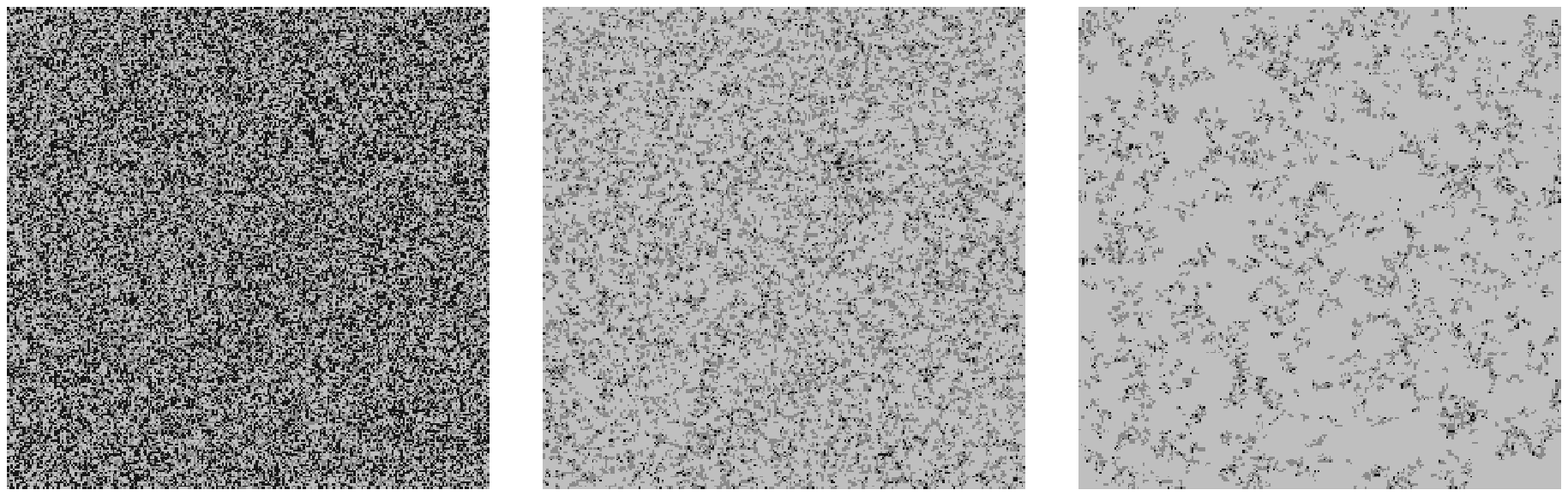}
\caption{Snapshots of the time evolution (time increases from left to right) of
the two-dimensional SLLVM model in the species coexistence phase, but near the 
predator extinction threshold, when the fixed point is a node. The light, gray,
and dark dots respectively represent the prey, predators, and empty sites on 
the lattice. The same rates, initial condition and system size apply as for 
Fig.~\ref{2d_snap}, except that now $\lambda = 0.15$.}
\label{2d_snap_node}
\end{figure*}

\subsection{Monte Carlo simulations of the SLLVM}

In this section, we report our results from direct Monte Carlo simulations in 
one, two, and three dimensions for the lattice reaction--diffusion (or 
stochastic lattice gas) SLLVM introduced in Sec.~III.A. 

The SLLVM under consideration is simulated on a simple cubic lattice with 
periodic boundary conditions. 
Each lattice site can be in one of the three possible states: occupied by a 
prey, by a predator, or empty. 
The algorithm that we use for simulating our model is the following:
\begin{itemize}
\item randomly choose a site on the lattice and generate a random number ($RN$)
      uniformly distributed between zero and one to perform the four possible 
      reactions of our SLLVM (with rates $D, \mu, \lambda$ and $\sigma$), 
      namely either of the four following processes:
\item if $RN < 1/4$ then randomly select one of the neighboring sites, and with
      rate $D$ exchange the contents of the two sites if the neighboring site 
      is empty;
\item if $1/4 \leq RN < 1/2$ and if the site holds a predator, then with rate
      $\mu$ the site will become empty;
\item if $1/2 \leq RN < 3/4$ and if the site holds a predator, choose a 
      neighboring site at random; if that site holds a prey then with rate 
      $\lambda$ the neighboring site becomes a predator;
\item if $3/4 \leq RN < 1$ and if the site holds a prey, randomly select a  
      neighboring site; if that site is empty, then with a rate of $\sigma$ the
      neighboring site becomes a prey.
\end{itemize}
One Monte Carlo step (MCS) is completed when the above steps are repeated as
many times as there are number of the sites on the lattice. 
We have numerically checked that explicit diffusion does not usually alter the 
behavior of the system, even when diffusion is fast compared to the reactions: 
we have run simulations with $D$ up to $1000$ times bigger than all the other 
rates, and not observed any qualitative changes. 
Specifically, we have verified that the spatial structures such as those 
depicted in Figs.~\ref{2d_snap} and \ref{2d_snap_node} were also obtained for 
small (or zero), intermediate, and  large ($D = 0 \ldots 1000$) diffusivities.
Similarly, the critical properties of the system (scaling exponents), as 
discussed in detail hereafter, were found to be independent of the values of 
the diffusion rate (at least in the range $D = 0 \ldots 1000$). 
Hence, without loss of generality and for the sake of simplicity, in many 
simulations we have set $D = 0$. 
Note that in this case, the particle offspring production processes 
effectively generate diffusive proliferation of the two species.

\begin{figure*}[t]
\includegraphics[angle=-90,width=6.5cm]{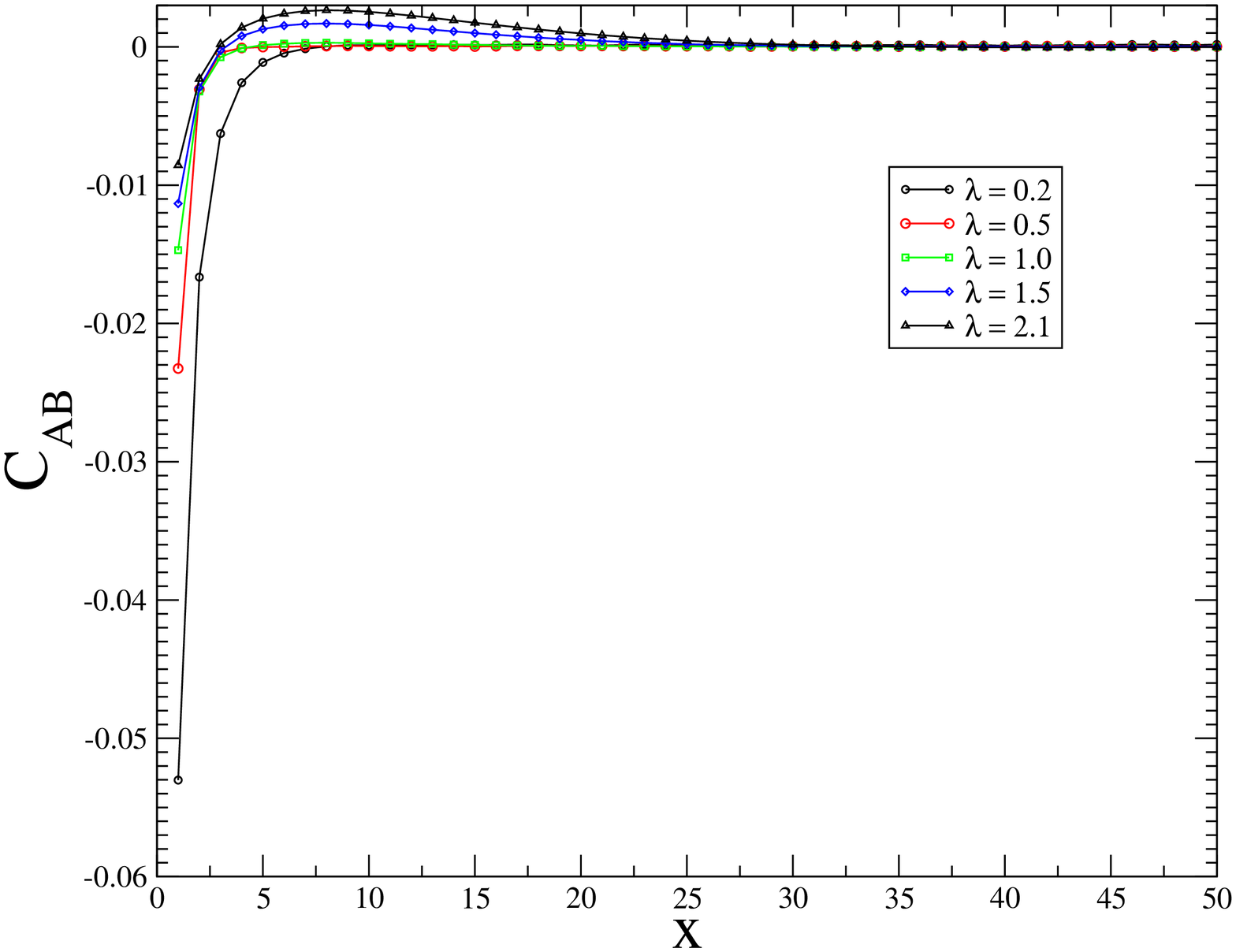}
\hspace{0.3cm}
\includegraphics[angle=-90,width=6.5cm]{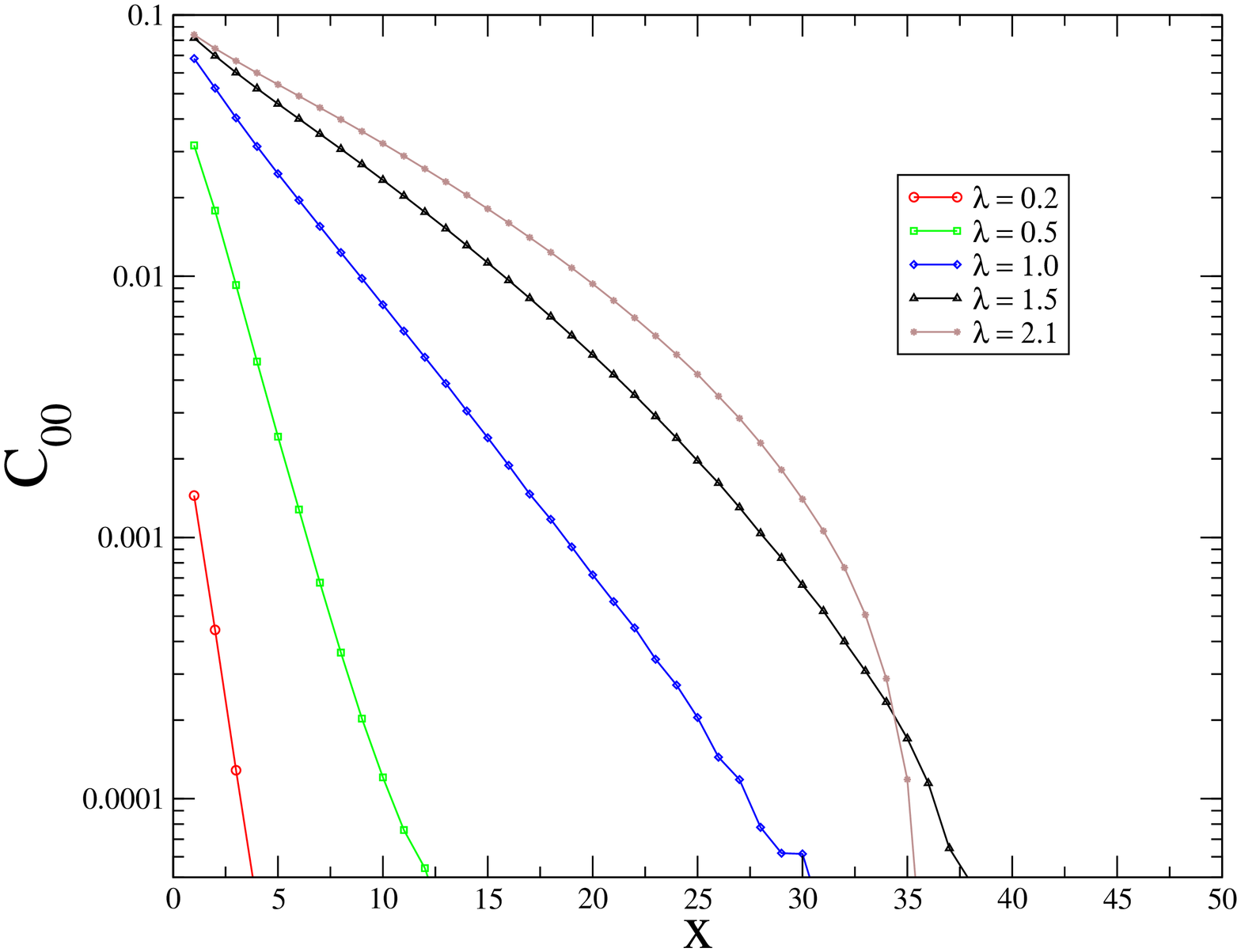}
\caption{(Color online.) Two-dimensional static correlation functions 
$C_{A,B}(x)$ (left) and $C_{\oslash,\oslash}(x)$ (right, in linear-logarithmic 
scale) for $\sigma = 4.0 $, $D = 0$, $\mu = 0.1$, and 
$\lambda = 0.2, 0.5, 1.0, 1.5, 2.1$. The system size is $256 \times 256$.}
\label{Cab}
\end{figure*}

Typical trajectories in the active phase (in 2D), all starting from random 
initial configurations, are depicted in Fig.~\ref{2d_traj}. 
In qualitative agreement with the mean-field analysis, the fixed point can 
either be a node, which in its vicinity is reached via straight trajectories, 
or, for larger values of the predation rate $\lambda$, a focus that is 
approached in spiralling paths. 
Of course, the agreement with the mean-field theory is {\em not} fully 
quantitative: 
in fact, there are fluctuations not accounted for in the rate equations
(\ref{EOMinter}), (\ref{EOMinter1}). 
However, the qualitative agreement (in dimensions $d > 1$) reported here on 
the structure of the phase portrait as predicted by the mean-field equations 
and as obtained for the SLLVM is remarkable, and actually was not observed 
in other stochastic predator--prey systems \cite{Provata,Lip}. 
For the various rates and the related values of the fixed points $a^*$ and 
$b^*$, we can also check that the inequalities (\ref{simplex}) are actually 
obeyed.

The three pictures on Fig.~\ref{2d_snap} show three consecutive snapshots of 
the system in the predator--prey coexistence phase on a two-dimensional 
lattice for parameter values for which the fixed point is a focus. 
We observe the formation of highly nontrivial patterns that display strong 
correlations between the predator and prey populations \cite{Movies1}. 
These typical snapshots illustrate how starting from a spatially homogeneous 
random initial configuration this simple model may develop amazingly rich 
patterns in the steady state where one can clearly distinguish fluctuating 
localized spots of predator and prey activity. 
In this regime we see that in the early stages of the system's temporal 
evolution rings of prey are formed that are followed by predators in the 
inner part of the rings (the leftmost picture in Fig.~\ref{2d_snap}). 
These rings subsequently grow with time and merge upon encounter. 
The steady state is maintained by a dynamical equilibrium of moving fronts 
of prey (with a typical length set by the value of the stochastic parameters) 
followed by predators that in turn leave behind empty sites that are needed 
for the next wave of prey to step in.

\begin{figure*}
\includegraphics[angle=-90,width=6.5cm]{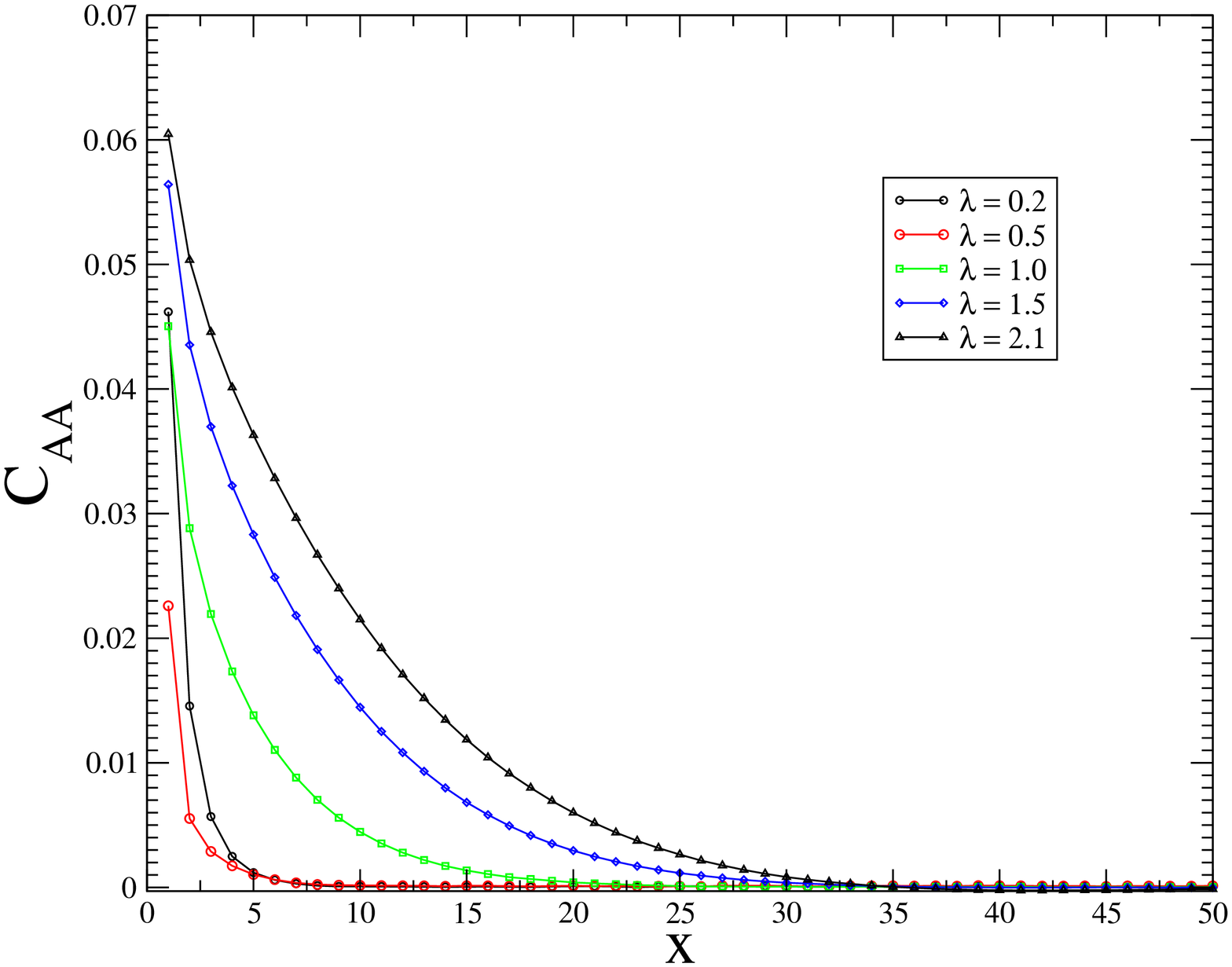}
\hspace{0.3cm}
\includegraphics[angle=-90,width=6.5cm]{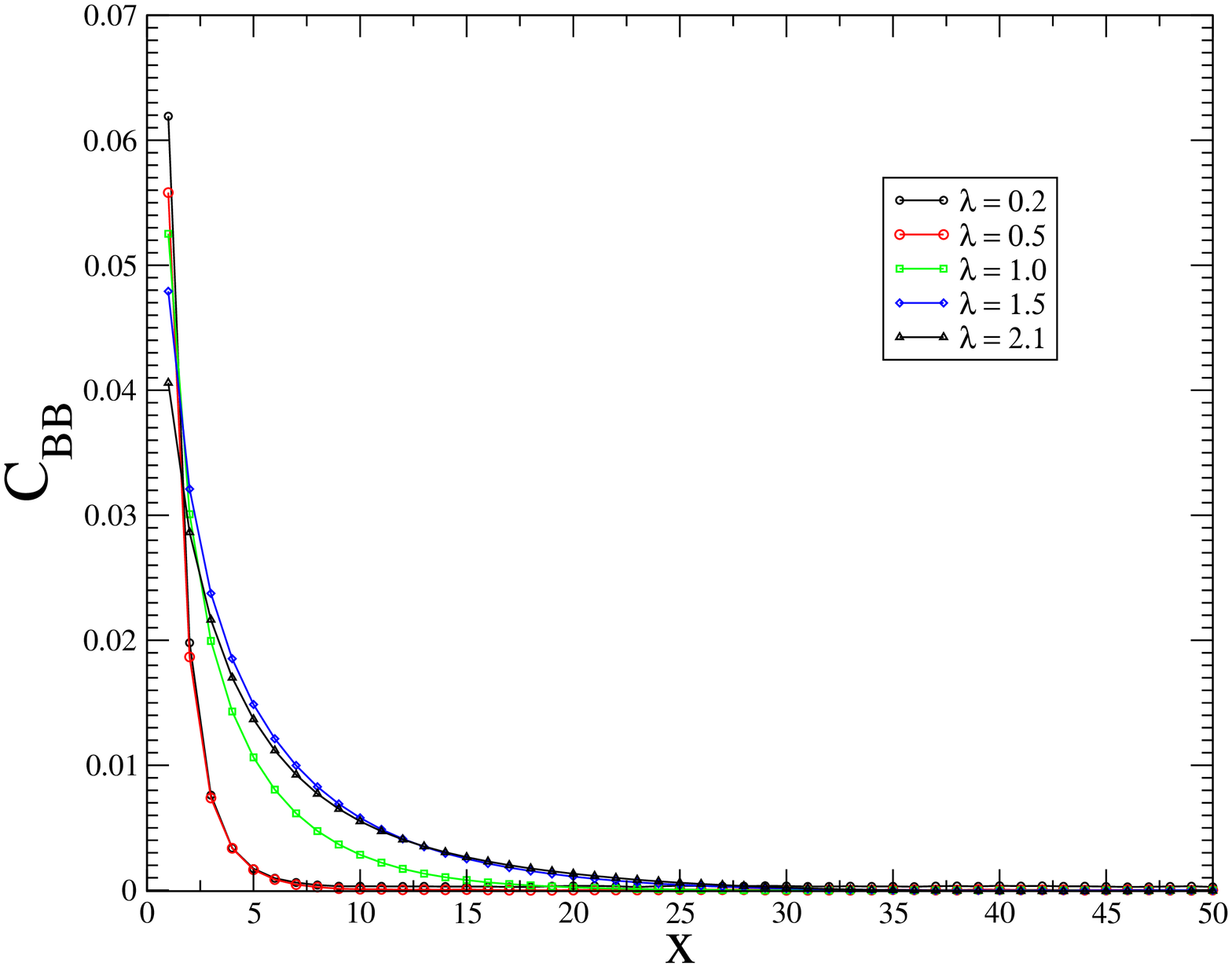}
\caption{(Color online.) Two-dimensional static correlation functions 
$C_{A,A}(x)$ (left) and $C_{B,B}(x)$ (right) for $\sigma = 4.0$, $D = 0$, 
$\mu = 0.1$ and $\lambda = 0.2, 0.5, 1.0, 1.5, 2.1$. 
The system size is $256 \times 256$.}
\label{Caa}
\end{figure*}

\begin{figure*}
\includegraphics[angle=-90,width=6.5cm]{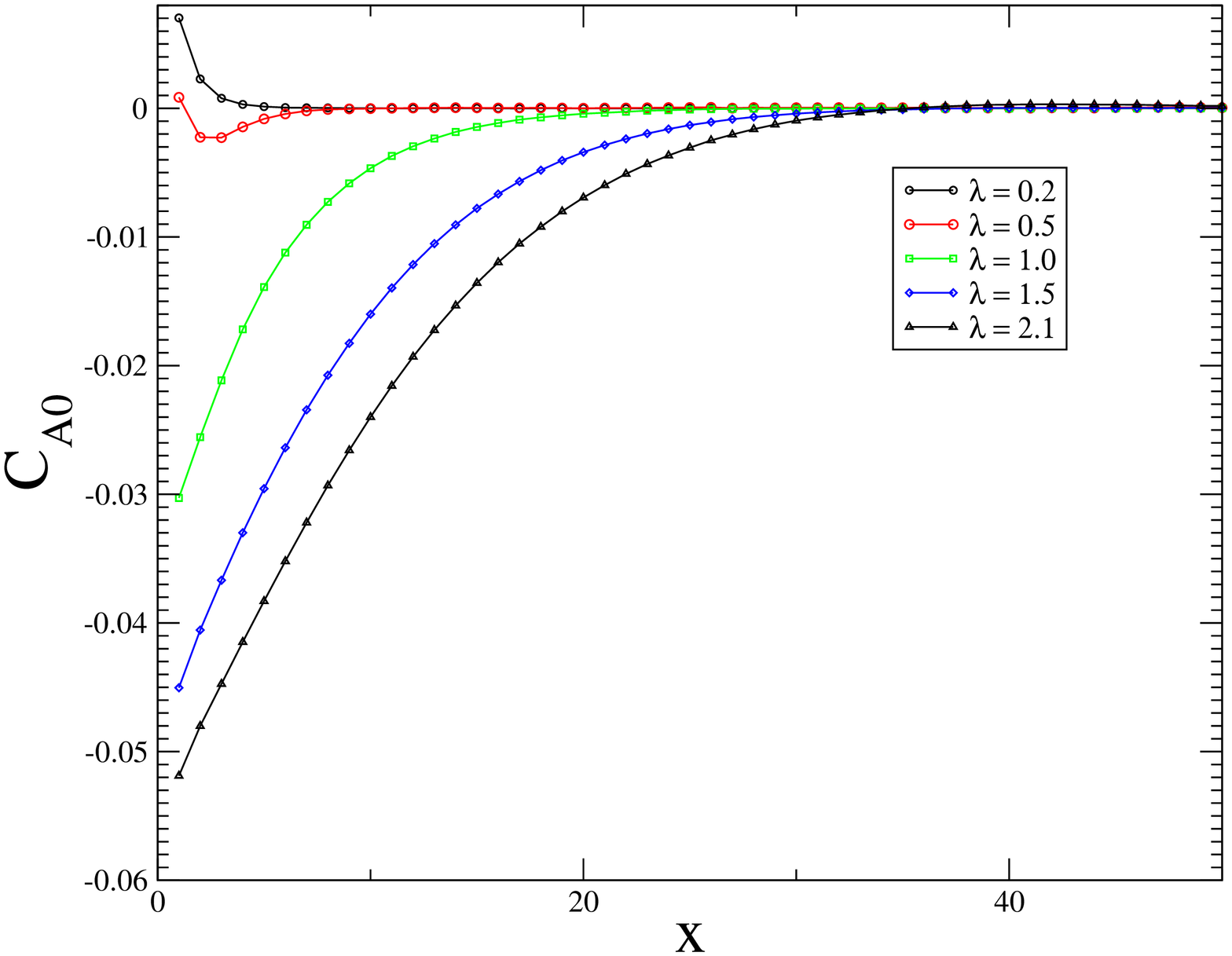}
\hspace{0.3cm}
\includegraphics[angle=-90,width=6.5cm]{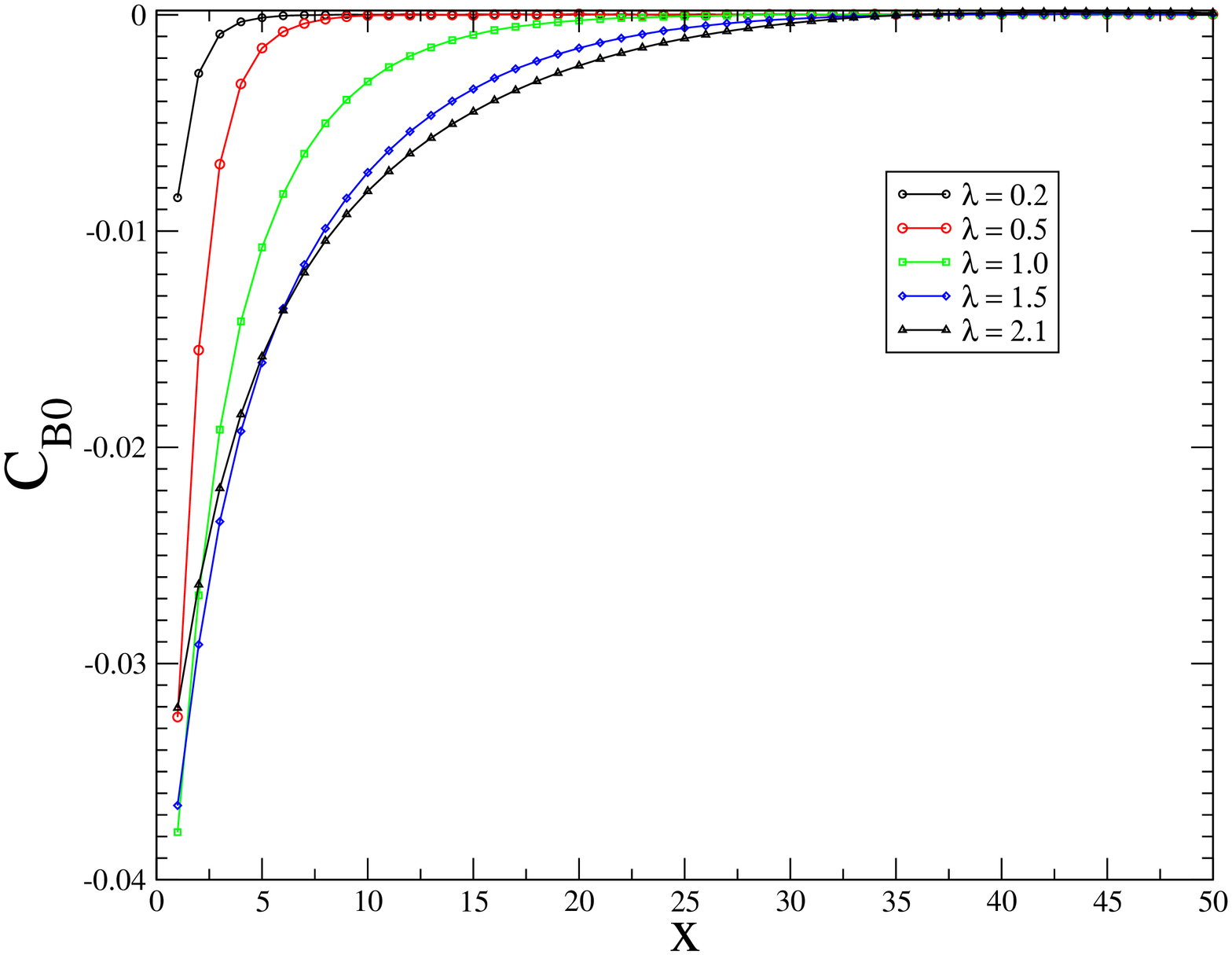}
\caption{(Color online.) Two-dimensional static correlation functions 
$C_{A,\oslash}(x)$ (left) and $C_{B,\oslash}(x)$ (right) for $\sigma = 4.0$, 
$D = 0$, $\mu = 0.1$ and $\lambda = 0.2, 0.5, 1.0, 1.5, 2.1$. 
The system size is $256 \times 256$.}
\label{Ca0}
\end{figure*}

To gain further quantitative insight on the complex spatial structure 
(see Fig.~\ref{2d_snap}) and on the fluctuations 
characterizing the system, we have computed numerically the static (and 
translationally invariant) correlation functions between various species, 
defined as $C_{\alpha,\beta}(x) = \langle n_{\bm j + \bm x}^\alpha \, 
n_{\bm j}^\beta \rangle(\infty) - \langle n_{\bm j + \bm x}^\alpha(\infty) 
\rangle \, \langle n_{\bm j }^\beta(\infty) \rangle$, where 
$\alpha,\beta \in (A,B,\oslash)$. 
For the sake of illustration, in Figs.~\ref{Cab}, \ref{Caa}, and \ref{Ca0} 
we report all the six connected correlation functions of the system, namely 
$C_{A,B}(x)$, $C_{\oslash,\oslash}(x)$, $C_{A,A}(x)$, $C_{B,B}(x)$, 
$C_{A,\oslash}$, and $C_{B,\oslash}$ measured for various two-dimensional 
situations.
The static correlation functions were obtained on $256 \times 256$ lattices
where the data were taken every $200$ MCS for a run of total $2\times 10^9$
MCS. 
When the rates $\sigma$, $D$, and $\mu$ are held fixed, the behaviors 
displayed by $C_{A,B}(x)$, $C_{\oslash,\oslash}(x)$, $C_{A,A}(x)$, 
$C_{B,B}(x)$, $C_{A,\oslash}$, and $C_{B,\oslash}$ can be qualitatively 
understood taking into account the fact that the predation reaction 
$AB \to AA$ occurs more likely when $\lambda$ is raised (as a consequence, 
the predators are more efficient in `chasing' the prey).
As shown in Fig.~\ref{Cab} (left), there is an effective repulsion at short
distances (anticorrelations for small $x$) and an effective attraction 
(positive correlations $C_{A,B}(x) > 0$) at larger (but finite) distances 
between predators and prey. 
This effective `attraction' results in the `bumps' (rounded peaks) of
 Fig.~\ref{Cab} for a relative distance of $x = 5-10$ lattice sites. 
These facts translate pictorially in the complex patterns displayed in
Fig.~\ref{2d_snap}, where the prey spots are typically at a finite distance 
from the predators: they are `eaten' if they come too close. 

Figures~\ref{Caa} and \ref{Ca0} show that (anti-)correlations 
[$C_{A,A}(x) > 0$, $C_{B,B}(x) > 0$, and $C_{A,\oslash}(x) < 0$, 
$C_{B,\oslash}(x) < 0$ at finite distance $x$] develop, respectively among 
predators, among prey, and between predators or prey and empty sites, when 
$\lambda$ is raised: 
predators (prey) effectively `attract' each other while the predators/prey 
and vacancies `repel' each other. 
In fact, in Fig.~\ref{2d_snap} we notice `clusters' of predators well 
separated from those of empty sites. 
Figure~\ref{Cab} (right) illustrates that correlations $C_{\oslash,\oslash}(x)$
among empty sites increases with the value of $\lambda$, which results from 
the `clustering' among predators and prey occurring in the coexistence phase,
as shown on the rightmost of Fig.~\ref{2d_snap} (see also \cite{Movies1}).

\begin{figure}[!t]
\includegraphics[angle=0,width=3.3in]{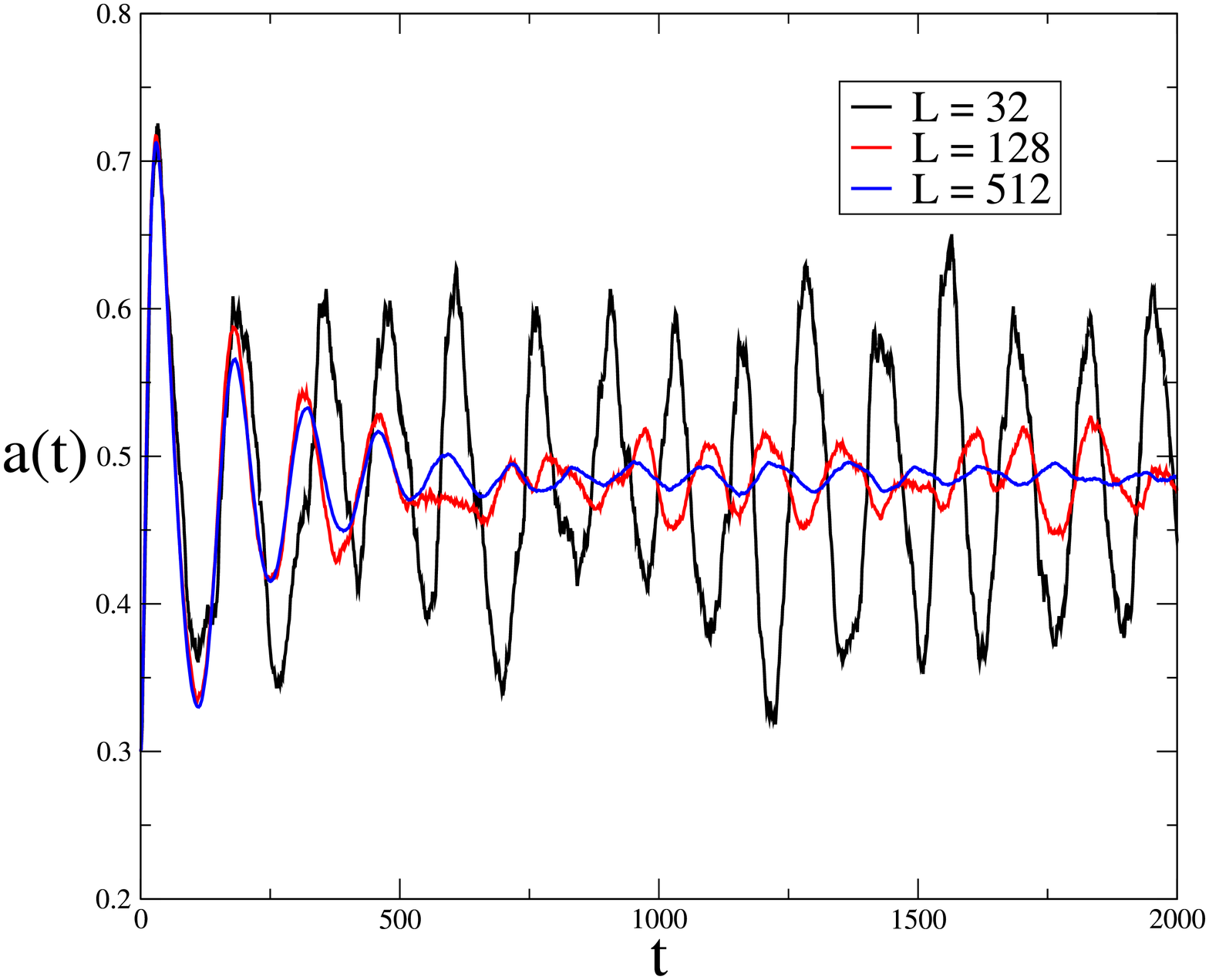}
\caption{(Color online.) The density of the predators $a(t)$ vs. $t$ on 
two-dimensional lattices (measured for single realizations) with 
$L = 32$, $128$ and $512$. 
The values of the stochastic parameters are $D = 0$, $\lambda = 1$, 
$\sigma = 4$, and $\mu = 0.1$. 
Initially the particles are homogeneously distributed with densities 
$a(0) = b(0) = 0.3$.}
\label{oscill_diff_L}
\end{figure}

\begin{figure}[!t]
\includegraphics[angle=0,width=3.3in]{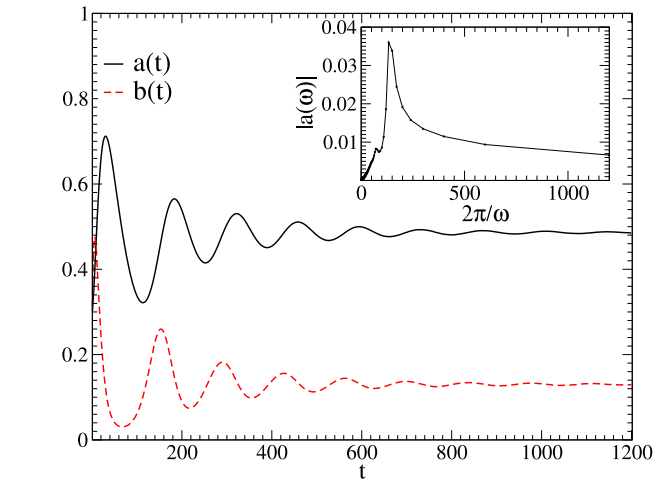}
\caption{(Color online.) The predator and prey densities $a(t)$ and $b(t)$ vs. 
$t$, from single runs, on a $4096 \times 4096$ lattice for $D = 0$, 
$\lambda = 1$, $\sigma = 4$, and $\mu = 0.1$. 
The inset shows the Fourier transform of the density of predators with a
pronounced peak at around $135 \times 5$ MCS.}
\label{L4096_fourier}
\end{figure}

Figure~\ref{oscill_diff_L} displays, for a single realization, i.e., without 
sample averaging), the typical temporal behavior of the predator density 
$a(t)$ when the fixed point is a focus. 
After some initial time interval of damped oscillations (see 
Fig.~\ref{L4096_fourier}), as consequence of the spatial fluctuations, the 
predator density oscillates in a rather erratic fashion around the average 
value. 
It is clear from the graphs that the amplitude of the oscillations in the 
steady state decreases with the system size; in the thermodynamic limit the
amplitude of the oscillations vanishes (see Fig.~\ref{L4096_fourier}). 
This remarkable feature was also reported in other stochastic lattice 
predator--prey systems \cite{Provata,Lip,TiborDroz}.

For our SLLVM variant, Fig.~\ref{L4096_fourier} depicts the transient 
regime (again, for a single realization) from a random starting configuration, 
initially filled with $1/3$ of each of the species, toward its steady state. 
The plots of the densities for both species exhibit damped oscillations with 
a period and amplitude that is completely independent of the initial 
conditions, in contrast with the predictions from the standard deterministic 
Lotka--Volterra rate equations (\ref{clotvol}), (\ref{blotvol}). 
The inset in Fig.~\ref{L4096_fourier} shows that the Fourier component 
$|a(\omega)|= | \sum_t e^{i\omega t} a(t)|$ vs. $2\pi / \omega$ displays a 
distinct peak at around $135 \times 5$ MCS (data are taken every $5$ MCS) for 
this set of values of the stochastic parameters, namely $D = 0$, $\lambda = 1$,
$\sigma = 4$, and $\mu = 0.1$.

\begin{figure}[!b]
\includegraphics[angle=0,width=3.3in]{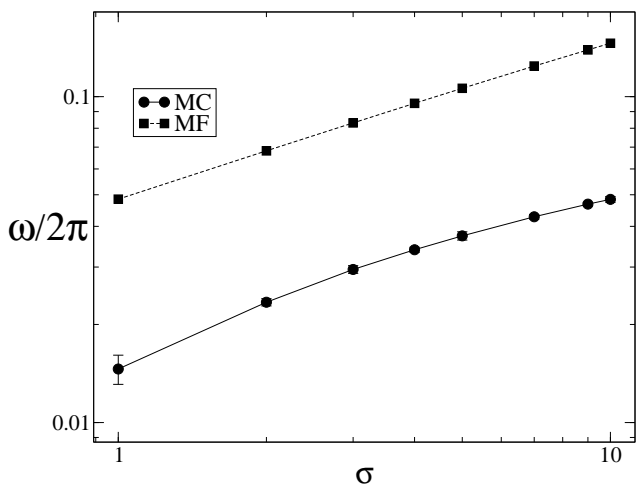}
\caption{Functional dependence of the characteristic frequency: comparison of 
mean-field prediction $\omega_{\rm MF}$ (squares) and Monte Carlo simulations 
(circles) on a $128 \times 128$ lattice. 
The reactions rates are $\lambda = 1.6$, $\mu = 0.1$, $D=0$.}
\label{freqdep}
\end{figure}

It is important to emphasize the fact that the amplitude of the oscillations 
decreases with increasing the lattice size only `globally', i.e., if one 
measures the total density of the species. 
In contrast, if one observes the temporal evolution of the density on a small, 
fixed subset of the lattice then the amplitude of the oscillations on this 
sub-lattice remains approximately the same with increasing the volume of the 
system \cite{Provata}.
The erratic oscillations displayed in (finite) predator--prey systems have
found considerable interest in the recent years. 
For instance, McKane and Newman \cite{McKane} have considered 
a zero-dimensional stochastic predator--prey model (represented as an `urn'), 
and have shown that the frequency predicted by the mean-field rate equations 
naturally appears to be the characteristic frequency of the damped 
oscillations of their model and results of a stochastic resonance 
amplification. 
We have also studied the functional dependence of the characteristic frequency 
$\omega_{\rm MC}$ of the damped erratic oscillations on the branching rate 
$\sigma$ (for fixed values of $\lambda, D$ and $\mu$) for a fairly large 
two-dimensional lattice (but still displaying some clear erratic behavior, see 
Fig.~\ref{oscill_diff_L}), and compared the results with the predicted 
$\omega_{\rm MF}$ arising from the mean-field theory, obtained (with 
$\rho = \zeta = 1$) from the imaginary part of Eq.(\ref{eigNTint}): 
$\omega_{\rm MF}=|\Im{\epsilon_\pm(a^*_3,b^*_3)}| = 
\frac{\mu \sigma}{2\lambda} \sqrt{1 - \frac{4\lambda}{\sigma} \, 
\left( 1 - \frac{\lambda}{\mu} \right)}$. 
The results are shown in Fig.~\ref{freqdep}. 
We see that the characteristic frequency $\omega_{\rm MC}$ of the SLLVM is 
always markedly {\em smaller} than the mean-field prediction 
$\omega_{\rm MF}$ (by a factor $2.5 \ldots 3$), but the functional dependence 
on the parameter $\sigma$ appears to be in fairly good agreement with the 
mean-field predictions.
Yet we note that for another stochastic predator--prey model variant, Antal 
and Droz reported completely different functional dependence for the 
mean-field and Monte Carlo results \cite{TiborDroz}.

\begin{figure*}
\includegraphics[angle=-90,width=12cm]{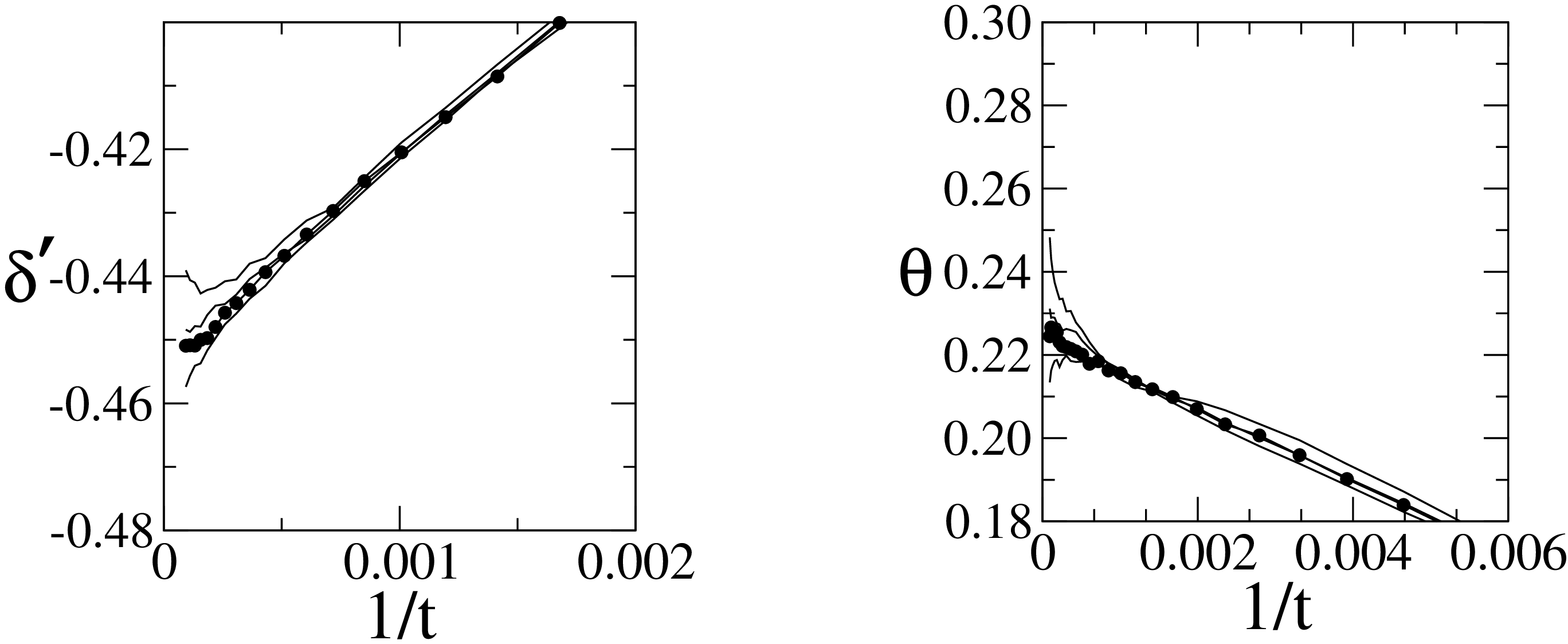}
\caption{Dynamical Monte Carlo simulations to estimate the predator extinction
threshold $\lambda_c$ for the two-dimensional (on a $512 \times 512$ lattice) 
NN model with $D = 0$, $\sigma = 4.0$, and $\mu = 0.1$. 
The effective scaling exponents $\delta'(t)$ vs. $1/t$ (on the left) and 
$\theta(t)$ vs. $1/t$ (on the right) are shown for four values of $\lambda$: 
$0.1690$, $0.1689$, $0.1688$, and $0.1687$ (from top to bottom).}
\label{delta_theta}
\end{figure*}

A completely different picture from that of Fig.~\ref{2d_snap} emerges when 
the fixed point is a node. 
The three plots in Fig.~\ref{2d_snap_node} again depict snapshots (starting 
with random initial configuration on the left) of the coexistence phase on a 
two-dimensional lattice \cite{Movies1}. 
In this case no real pattern formation takes place in the steady state; 
rather we notice a small number of `clouds' (clusters) of predators 
effectively diffusing in a sea of prey (the rightmost picture on 
Fig.~\ref{2d_snap_node}). 
Upon lowering the value of $\lambda$ further, the average size of the 
predator `clouds' and their density decreases, and the system eventually 
enters the absorbing phase for sufficiently small values of $\lambda$. 
One observes here that the dynamics of the small activity clusters close to 
the absorbing transition is very simple:
(i) an `active spot' can die; (ii) upon encounter two (or more) `activities' 
usually coalesce; (iii) an `active spot' can split into two (branching). 
Thus, as the system displays a continuous phase transition from a fluctuating 
active phase into a unique stable absorbing state; as only short-range 
interactions are involved; and since the model is not subject to any special 
symmetries or conservation laws, the conditions of the so-called DP 
conjecture \cite{DPconjecture} are fulfilled. 
Therefore, the phase transition occurring in this model from an active to 
the absorbing phase (from the predators' viewpoint) is a good candidate for 
the directed percolation (DP) universality class \cite{DPreviews,Howard}. 

For studying the critical properties of the model for the transition from an 
active to an absorbing state, we employ as an order parameter the average 
predator density $a$. 
In the active phase, $a(t \to \infty)=a^*$ assumes a nonzero value, while in 
the absorbing state the lattice is full of prey and $a^* = 0$.
Thus, at the critical point, the exponent $\beta$ is defined according to
$a^*\sim (\lambda-\lambda_c)^{\beta}$ \cite{DPreviews}.
The stochastic fluctuations are responsible for a shift of the critical 
point and changes in the critical exponents: 
for instance (in dimensions $1 < d \leq 4$), the actual computed value of 
$\beta$ is always smaller than the value $\beta_{\rm MF} = 1$ predicted by 
the mean-field analysis. 

In order to check the critical properties of the two-dimensional model close 
to the extinction phase transition point, we employ the dynamical Monte Carlo 
approach with an initial configuration that has only a single active site (a 
predator) in the middle of the lattice, with the remainder filled with prey 
\cite{DPreviews}. 
As an illustration, in Fig.~\ref{delta_theta} we report the dynamical Monte 
Carlo analysis for a $512 \times 512$ lattice with $D = 0$, $\sigma = 4.0$, 
and $\mu = 0.1$. 
In this case, the duration of the simulations was $10^5$ MCS. 
We chose to measure the survival probability $P(t)$ [the probability that 
after time $t$ we still have predators in the system], and the number of 
active sites (i.e. predators) $N(t)$. 
In order to obtain reasonably good estimates for these two quantities, we 
performed $3 \times 10^6$ independent runs. 
Close to the critical point, $P(t)$ and $N(t)$ follow algebraic power laws 
with critical exponents $\delta$ and $\theta$, respectively:
\begin{equation}
P(t) \sim t^{-\delta'} \ , \qquad N(t) \sim t^{\theta} \ .
\end{equation}
Figure~\ref{delta_theta} shows the effective exponents $\delta'(t)$ and 
$\theta(t)$ defined via
\begin{eqnarray}
\label{delta_thet}
- \delta'(t) &=& \frac{\ln[P(t)/P(t/3)]}{\ln 3} \ , \nonumber \\
\theta(t) &=& \frac{\ln[N(t)/N(t/3)]}{\ln 3} \ .
\end{eqnarray}
Just below (above) $\lambda_c$ the effective exponent graphs are supposed to 
curve down (up) for large $t$ while for $\lambda = \lambda_c$ they should be 
more or less straight lines; their intercept gives the numerical value of the
exponent. 
From the graphs we estimate the critical value of $\lambda_c = 0.1688(1)$ 
instead of the mean-field result $\lambda_c = \mu = 0.1$; as to be expected, 
fluctuations shift the critical point to larger values of the predation rate 
(suppress the `ordered' phase).
The values for $\delta'$ and $\theta$ are very close to $0.451$ and $0.230$, 
respectively, which are the known exponents for the two-dimensional DP model
\cite{DPreviews}. 
For the system depicted in Fig.~\ref{delta_theta}, the numerical value of the 
critical exponent $\beta$ is also very close to the established 
$\beta \approx 0.584$ exponent for the two-dimensional DP model. 
We have checked that for other choices of the rates $D$, $\mu$, and $\sigma$ 
we also obtain critical exponents that are consistent with the DP universality 
class.

The Monte Carlo simulations for the three-dimensional model result in values 
for the critical exponents that are again very close to the established DP 
critical values. 
For instance, near the critical point, we have measured an exponent 
$\beta \approx 0.81$, in excellent agreement with the corresponding value, 
$\beta_{\rm DP}\approx 0.81(1)$ reported for DP in $d = 3$ \cite{DPreviews}. 
In three dimensions, we also observe the same two different scenarios, namely
isolated predator clusters near the threshold and expanding and merging 
activity fronts at larger predation rates, as in two dimensions, see
Figs.~\ref{2d_snap} and \ref{2d_snap_node}). 
Not surprisingly, we have found that the complex patterns associated with the
active focus fixed point are less correlated in $d = 3$ compared with $d = 2$. 
Also, for dimensions $d > 4$ we recover the mean-field critical exponents, 
consistent with the fact that the upper critical dimension is $d_c = 4$ for 
the DP universality class.

Numerical results suggesting that lattice predator-prey models exhibit an 
active-absorbing phase transition belonging to the DP universality class have
also been reported recently for other two-dimensional model systems 
\cite{Lip,TiborDroz,Albano}. 
In Sec.~III.D, we provide field-theoretic arguments that support the assertion
that the critical properties near the predator extinction threshold in these 
models is indeed generically described by the DP scaling exponents (see also
Ref.~\cite{Janssen}).
We have also performed Monte Carlo simulations for systems where the predation
reactions were subject to a spatial bias, i.e., possible only along a special
direction in two dimensions.
While such a bias clearly renders the activity fronts in the active phase
anisotropic, it does not seem to affect the properties near the extinction
threshold.
For aside from an overall slow drift along the preferred spatial direction, 
which sets up a net particle current, the predators still form isolated 
islands in a sea of prey.
Hence we expect that one should observe the DP critical exponents even in this
`driven' system, see Sec.~III.D below.
Similarly, when the predators are made to actually `follow' the prey, by 
biasing the hopping probabilities for the $A$ species towards neighboring 
sites occupied by $B$ particles, no qualitative changes from the simple SLLVM
are observed. 
Notice that this variant of the SLLVM differs from that considered in 
Refs.~\cite{Albano}: whereas there both predators and prey were allowed to 
perform `smart moves', we have allowed only the predators to `chase' the prey 
by moving toward the regions where the concentration of prey is locally 
highest. 
These differences might be important as they could perhaps explain that we 
always observed damped erratic oscillations, whilst the authors of 
Refs.~\cite{Albano} reported the existence (in $d = 2$) of self-sustained 
oscillations.

\begin{figure}[!t]
\includegraphics[width=2.6in]{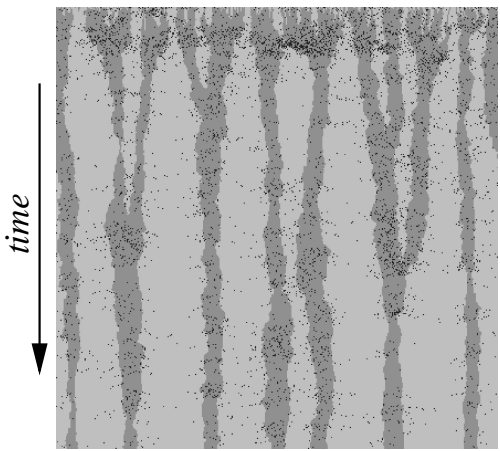}
\caption{Space-time plot for the one-dimensional SLLVM (with mutual 
predator--prey site restrictions), starting from a random initial 
configuration with homogeneous density distribution,  $a(0)=0.5 = b(0)$ (top 
row; time proceeds downwards).   
Multi-strip metastable configurations are observed for a long time while the
system slowly evolves toward the absorbing steady state devoid of predators.
The system size here is $L=1024$, and for the parameters we chose $D=1$, 
$\mu=0.0005$, $\sigma=0.01$, and $\lambda=0.008$. 
The light gray, dark gray, and black dots respectively represent the prey, 
predators, and empty sites.}
\label{1d_stripes}
\end{figure}

To conclude this section, we briefly consider the one-dimensional case.
It is well-known that site restrictions may become quite crucial in one
spatial dimension \cite{RD,Odor,DHT}, one must be prepared to encounter 
special behavior in this case. 
Indeed, since in our version of the SLLVM we are not allowing simultaneous
site occupation by the predators and prey, in contrast with 
Refs.~\cite{Lip,TiborDroz}, the $A$ and $B$ populations may be forced to
segregate into distinct domains on a one-dimensional lattice (see the similar
mechanisms for other multi-species reaction--diffusion systems reported in 
Refs.~\cite{Odor,DHT}).
The results of our computer simulations in fact show that the steady state of 
the one-dimensional system is a lattice full of prey for {\em all} values of 
the stochastic parameters. 
If we start from a random initial configuration we observe that the system 
`coarsens' (as in the one-dimensional three-species cyclic Lotka--Volterra 
model, see Ref.~\cite{cyclic} and the end of Sec.~I): 
it slowly evolves into configurations of repeating sequences of domains of 
predators and prey. 
The spatio-temporal plot of a typical run is shown in Fig.~\ref{1d_stripes}.
This multi-domain configuration constitutes a long-lived metastable state in 
the one-dimensional system, and typically an enormous crossover time must 
elapse for the system to reach the steady state, even in our finite lattices. 
As time increases,  domains (`stripes') of predators merge in a
sea of prey (with some sparse holes) and eventually, in the steady state, the
number of these stripes of predators vanishes for any set of values of the 
stochastic parameters; the final system is full of prey. 

We have indeed found the width of a single predator--hole domain to remain 
constant upon increasing the system size, which is typical of a coarsening 
phenomenon, and supports the previous observation that asymptotically, in the 
thermodynamic limit, one arrives at a steady state with vanishing predator 
density. 
We might think of the effective long-time coarse-grained dynamics of the 
predator and prey domains as being described by the simple coagulation/decay
reactions $\tilde A + \tilde A \to \tilde A$ and $\tilde A \to \emptyset$.
where $\tilde A$ represents a predator--hole domain, and $\emptyset$ indicates
a prey domain. 
As $t \to \infty$, this would suggest that the predator density should decay
as a power law $\sim t^{-1/2}$, ultimately turning over to an exponential
cutoff. 
Owing to the huge crossover times in this system, we were not able to confirm
this conjecture quantitatively.
Yet the same kind of behavior was reported in Ref.~\cite{Provata} for the 
one-dimensional version of the cyclic three-state SLLVM.
Notice, however, that since in the four-state models of 
Refs.~\cite{Lip,TiborDroz} predator and prey particles were allowed to occupy
the same lattice sites, species segregation did not occur in the simulations
reported there.

\subsection{Field-theoretic analysis of the continuous predator extinction 
	transition}

We have seen that our Monte Carlo simulations in two and three dimensions in
many ways confirm the {\em qualitative} picture from the mean-field
predictions, once growth-limiting terms are taken into account there.
However, the mean-field approximation naturally cannot capture the intriguing 
dynamical spatial structures in the active coexistence phase of the SLLVM.
Moreover, it does not aptly describe the universal scaling properties near the
predator extinction threshold.
In the language of nonequilibrium statistical mechanics, and with respect to
the predator population, this constitutes a continuous phase transition from an
active phase to an inactive, absorbing state.
For local interactions, and in the absence of additional conservation laws and
quenched disorder, such active to absorbing state phase transitions are known
to be generically described by the universality class of directed percolation
(DP) \cite{DPconjecture} (for recent reviews, see Refs.~\cite{DPreviews}).
Since this is true even for many-species systems \cite{Janssen}, one would
expect the extinction threshold in the SLLVM to be governed by the DP 
exponents as well, as was indeed suggested in Refs.~\cite{Tome,Lip,TiborDroz}.
In the preceding Sec.~III.C, we have added further evidence from our Monte
Carlo simulation data that the critical exponents in the SLLVM are consistent
with those of DP.

We now proceed to provide field-theoretic arguments, based on a standard 
mapping of the master equation corresponding to the SLLVM processes, i.e., the 
spatial extension of Eq.~(\ref{0dstolv}). 
For the sake of clarity, before providing our field-theoretic analysis of the 
SLLVM, we briefly outline the general approach (see Ref.~\cite{Howard} and 
references therein). 
The first step towards a field-theoretic treatment is to recast the master 
equation (\ref{master}) into the stochastic (quasi-)Hamiltonian formulation 
\cite{RD}, the site restrictions being implemented into a second quantization 
bosonic formalism (to avoid a more cumbersome representation in terms of spin 
operators) \cite{fred}. 
One then proceeds by adopting a coherent-state path integral representation and
by taking the continuum limit. 
This leads to an Euclidean action $S$ which embodies the statistical weight of 
each possible configuration and is the central ingredient to perform 
perturbative renormalization group (RG) calculations (the bilinear part of the 
action being identified as the Gaussian reference action). 
Here, our goal is to show that the action of the SLLVM can be mapped onto a 
field theory known to share the same critical properties as the DP. 
In addition, the field-theoretic treatment allows us to systematically 
discriminate between processes which are relevant/irrelevant for the critical 
properties of the SLLVM (in the RG sense) and thus identify robust features of 
the stochastic lattice predator-prey systems.

At this point, we specifically turn to field-theoretic analysis of the SLLVM.
The essence of the following treatment is  simply the observation that
the prey population is nearly homogeneous and constant near the predator 
extinction threshold.
The processes involving the prey then effectively decouple, and the SLLVM 
reactions essentially reduce to $A \to \oslash$ and $A \to A + A$.
Yet this set of processes, supplemented with either the growth-limiting 
reaction $A + A \to A$ or site restrictions for the $A$ particles, is just 
prototypical for DP \cite{DPreviews}.
Following the aforementioned standard procedures \cite{Howard}, the field 
theory action for the combined processes $A \to \oslash$ (rate $\mu$), 
$B \to B + B$ (rate $\sigma$), and $A + B \to A + A$ (rate $\lambda$), along 
with particle diffusion, becomes (omitting temporal boundary terms):
\begin{eqnarray}
\label{lvft}
  S[{\hat a},a;{\hat b},b] &=& \int \! d^dx \! \int \! dt \biggl[ 
  {\hat a} \left( \partial_t - D_A \, \nabla^2 \right) a \\ 
  &&\qquad + {\hat b} \left( \partial_t - D_B \, \nabla^2 \right) b + 
  \mu \left( {\hat a} - 1 \right) a \nonumber \\
  &&+ \sigma \left( 1 - {\hat b} \right) {\hat b} \, b \, 
  e^{- \rho^{-1} \, {\hat b} \, b} + \lambda 
  \left( {\hat b} - {\hat a} \right) {\hat a} \, a \, b \biggr] \ . \nonumber
\end{eqnarray}
Since we are interested in the critical properties near the extinction 
threshold, where there remains almost no predators in the system, we do not need
to take into account any spatial restriction imposed on the prey species by
the $A$ particles, and therefore have set $\zeta \to \infty$ in 
Eq.~(\ref{lvft}).
Note that $D_A$ and $D_B$ here are the {\em effective} diffusivities for the 
two species that in the site-restricted system emerge as a consequence of the
offspring production on neighboring sites (even if there is no hopping present
in the microscopic model).
The fields ${\hat a}$ (${\hat b}$) and $a$ ($b$) originate from the 
coherent-state left and right eigenvalues of the bosonic creation and 
annihilation operators in the stochastic (quasi-)Hamiltonian for the predators 
(prey).
The Trotter formula combined with the discrete hopping processes yields the
diffusion propagators in the action (\ref{lvft}), while the reactions are
encoded in the terms proportional to the rates $\mu$, $\sigma$, and $\lambda$.
In each of these terms, the first contribution indicates the `order' of the
reaction (namely, which power of the particle densities ${\hat a} \, a$ and
${\hat b} \, b$ enters the rate equations), whereas the second contribution
directly encodes the process under consideration ($a$ and $b$ annihilate a
predator or prey particle, ${\hat a}$ and ${\hat b}$ create them).

The exponential in the prey reproduction term captures the site restrictions
within the bosonic field theory \cite{fred}; the parameter $\rho$, with 
dimension of a particle density, emerges upon taking the continuum limit.
In terms of an arbitrary momentum scale $\kappa$, the scaling dimension of
$\rho^{-1}$ is therefore $\kappa^{-d}$, and it constitutes an irrelevant 
coupling in the renormalization group sense.
However, since it is essential for the existence of the phase transition,
we may not set $\rho^{-1} = 0$ outright, despite the fact that it does scale 
to zero under repeated scale transformations.
Rather, we may expand 
$e^{- \rho^{-1} \, {\hat b} \, b} \approx 1 - \rho^{-1} \, {\hat b} \, b$, 
but need to retain the first-order contribution.
The classical field equations $\delta S / \delta a = 0 = \delta S / \delta b$
are solved by ${\hat a} = 1 = {\hat b}$ (as a consequence of probability
conservation \cite{Howard}), whence 
$\delta S / \delta {\hat a} = 0 = \delta S / \delta {\hat b}$ then essentially
yield the mean-field equations of motion (\ref{dlotvol}) for $\zeta^{-1} = 0$,
and we may identify $\rho^{-1}$ with the prey carrying capacity.
Upon shifting the fields ${\hat a} = 1 + {\tilde a}$, 
${\hat b} = 1 + {\tilde b}$, the action then reads
\begin{eqnarray}
\label{lvfts}
  &&S[{\tilde a},a;{\tilde b},b] = \int \! d^dx \! \int \! dt \biggl[ 
  {\tilde a} \left( \partial_t - D_A \, \nabla^2 + \mu \right) a \nonumber \\ 
  &&\qquad\qquad\qquad + {\tilde b} \left( \partial_t - D_B \, \nabla^2 - 
  \sigma \right) b - \sigma \, {\tilde b}^2 \, b \nonumber \\
  &&\quad + \sigma \, \rho^{-1} \, (1 + {\tilde b})^2 \,
  {\tilde b} \, b^2
  - \lambda \, (1 + {\tilde a}) \, ({\tilde a} - {\tilde b}) \, a \, 
  b \biggr] \ . \quad
\end{eqnarray}

For vanishing predation rate $\lambda = 0$, the $A$ and $B$ processes of
course decouple, with the predators dying out (with rate $\mu$), whereas the
prey population is at its carrying capacity, $b \approx \rho$.
We are interested in the properties near the predator extinction threshold at
$\lambda_c$ (in mean-field theory, $\lambda_c = \mu / \rho$), where also
$a \to 0$ and $b \to b_s \approx \rho$.
We therefore introduce the fluctuating field $c = b_s - b$, and demand that
$\langle c \rangle = 0$, which eliminates the linear source term 
$\sim b_s \, {\tilde c}$ in the ensuing action.
With ${\tilde c} = - {\tilde b}$, we obtain
\begin{eqnarray}
\label{lvftc}
  &&S[{\tilde a},a;{\tilde c},c] = \int \! d^dx \! \int \! dt \, \bigl[ 
  {\tilde a} \left( \partial_t - D_A \, \nabla^2 + \mu - \lambda \, b_s 
  \right) a \nonumber \\ 
  &&+ {\tilde c} \left[ \partial_t - D_B \nabla^2 
  + (2 b_s / \rho - 1) \, \sigma \right] c 
  + \sigma \, b_s (2 b_s / \rho - 1) \, {\tilde c}^2 \nonumber \\
  &&- \sigma \, \rho^{-1} \, b_s^2 \, {\tilde c}^3 
  - \sigma \, (4 b_s / \rho - 1) \, {\tilde c}^2 \, c 
  - \sigma \, \rho^{-1} \, (1 + {\tilde c}^2) \, {\tilde c} \, c^2 \nonumber\\
  &&+ 2 \sigma \, \rho^{-1} \, {\tilde c}^2 \, (c + b_s \, {\tilde c}) \, c 
  - \lambda \, b_s \left[ {\tilde a}^2 + (1 + {\tilde a}) \, {\tilde c} 
  \right] a \nonumber \\
  &&+ \lambda \left( 1 + {\tilde a} \right) 
  \left( {\tilde a} + {\tilde c} \right) a \, c \bigr] \ .
\end{eqnarray}

We now exploit the fact that the prey density is hardly fluctuating, as 
encoded in the mass term $\approx \sigma$ for the ${\tilde c} \, c$ propagator.
Thus upon rescaling $\phi = \sqrt{\sigma} \, c$ and 
${\tilde \phi} = \sqrt{\sigma} \, {\tilde c}$, and letting $\sigma \to \infty$
(the scaling dimension of the branching rate $\sigma$ is $\kappa^2$, whence it
constitutes a relevant variable that will flow to infinity under the RG), the
nonlinear terms in the prey fields disappear, and the predator and prey 
sectors effectively decouple,
\begin{eqnarray}
\label{lvfta}
  S_\infty[{\tilde a},a;{\tilde \phi},\phi] &=& \int \! d^dx \! \int \! dt \, 
  \bigl[ {\tilde a} \left( \partial_t - D_A \, \nabla^2 
  + \mu - \lambda \, b_s \right) a \nonumber \\ 
  &&\qquad\quad - \lambda \, b_s \, {\tilde a}^2 \, a 
  + {\tilde \phi} \, \phi + b_s \,  {\tilde \phi}^2 \bigr] \ .
\end{eqnarray}
The fields $\phi$ and ${\tilde \phi}$ are now readily integrated out; for the
predators, however, we need to implement a growth-limiting term as originally
enforced through the finite supply of prey.
This is done most easily through adding the reaction $A + A \to A$ with rate 
$\tau$.
In the field theory action, this leads to the additional terms 
$\tau \left( {\hat a} - 1 \right) {\hat a} \, a^2 = 
\tau \, {\tilde a} \left( 1 + {\tilde a} \right) a^2$.
Setting $D_A \, r_A = \mu - \lambda \, b_s$, and rescaling the fields to
${\cal S} = \sqrt{\lambda \, b_s / \tau} \, a$ and 
${\widetilde {\cal S}} = \sqrt{\tau / \lambda \, b_s} \, {\tilde a}$, we finally
arrive at
\begin{eqnarray}
\label{lvftf}
  S_\infty[{\widetilde {\cal S}},{\cal S}] &=& \int \! d^dx \! \int \! dt \, 
  \Bigl[ {\widetilde {\cal S}} \Bigl( \partial_t + D_A \, (r_A - \nabla^2) \Bigr) {\cal S} 
  \nonumber \\ 
  &&\qquad\quad - u \, {\widetilde {\cal S}} \left( {\widetilde {\cal S}} - {\cal S} \right) {\cal S} 
  + \tau \, {\widetilde {\cal S}}^2 \, {\cal S}^2 \Bigr] \ , \quad
\end{eqnarray}
where $u = \sqrt{\tau \, \lambda \, b_s}$.
Since the scaling dimensions of both $\lambda$ and $\tau$ are $\kappa^{2-d}$,
and $b_s$ represents a particle density, the scaling dimension of the new
effective nonlinear coupling $u^2$ is $\kappa^{d-4}$.
Thus the upper critical dimension of the effective field theory (\ref{lvftf})
for the predator extinction threshold is $d_c = 4$, and the four-point vertex
$\propto \tau$ is irrelevant in the RG sense near $d_c$.
If we are interested only in asymptotic universal properties, we may thus drop
this vertex, which leaves us precisely with Reggeon field theory that 
describes the critical properties of DP clusters \cite{DP,DPreviews}.
Notice, however, that the above treatment does not apply to one dimension, if
we consider hard-core particles or site exclusion; as observed in other
reaction--diffusion models as well \cite{Odor,DHT}, the one-dimensional 
topology then induces species segregation, and in the system under 
consideration here, the active coexistence state disappears entirely.

We remark that the above processes also generate $A + B \to A$,
$A + B \to \oslash$, $B \to B + B + B$ etc. on a coarse-grained level.
These reactions can be readily included in the previous analysis, without any
qualitative changes in the final outcome: 
the continuous active to absorbing state phase transition for the predator 
population $A$ is in any case described by the DP critical exponents.
We had previously mentioned another variant, where the predation process is
spatially biased along a given direction.
If we mimic such a situation by an additional vertex of the form 
$- E \nabla_\parallel S^2$ (where the spatial derivative $\nabla_\parallel$ is 
along the drive direction), which is characteristic of driven diffusive 
systems \cite{DDS}, we note that since its scaling dimension is 
$[E] = \kappa^{2-d}$, it should be irrelevant at the extinction threshold, and 
the critical exponents of the phase transition still be described by DP.
(This argument assumes, however, that the drive nonlinearity does {\em not}
induce an anisotropy in the spatial ordering; if only the spatial sector
transverse to the drive softens, while the longitudinal fluctuations remain
noncritical, novel critical behavior may ensue.)

\section{Conclusion}

We have studied the effect of spatial constraints and stochastic noise on the 
properties of the Lotka--Volterra model, which is a generic two-species
predator--prey system defined on a lattice interacting via a predation 
reaction that involves nearest neighbors.
We obtain a rich collection of results that differ remarkably from the
predictions of the classical (unrestricted) deterministic Lotka--Volterra 
model.
This investigation was carried out both analytically, using a suitable 
mean-field approach and field-theoretic arguments, as well numerically, 
employing Monte Carlo simulations. 
(In this paper, we have mainly presented figures obtained from two-dimensional
simulations, the most ecologically relevant situation, but we have also 
checked our statements running simulations in one, three, and four 
dimensions.)

The mean-field analysis of the stochastic lattice Lotka-Volterra model (SLLVM)
\cite{Murray} predicts that there is a continuous non-equilibrium phase 
transition from an active (species coexistence) to an absorbing (full of prey)
state and these predictions are confirmed, in dimensions $d>1$, by the 
computer simulations. 
Already in other stochastic lattice predator--prey models, it was shown that 
the mean-field description provided this type of behavior (see, e.g., 
Refs.~\cite{Tome,TiborDroz,Albano}). 
From an ecological and biological perspective, this means that the rate 
equations, when they take into account limited local resources, predict a
possible extinction of one population species (here, the predators), which is 
a realistic feature absent from the conventional Lotka--Volterra rate 
equations \cite{Murray}.

Actually, in contrast to the cyclic three-state SLLVM of Ref.~\cite{Provata} 
and other stochastic predator-prey models \cite{Lip}, here the mean-field 
predictions capture the essential qualitative features of the SLLVM phase
diagram in dimensions $d > 1$.
Our field-theoretic analysis shows that the mean-field critical exponents 
are quantitatively valid in dimensions $d > d_c = 4$.
According to the mean-field analysis and the Monte Carlo simulations, the 
stable coexistence fixed points of this model can be either nodes or foci. 
It does not exhibit stable and persistent population oscillations but only 
damped and transient ones, near a stable focus, whose amplitudes vanish in 
the thermodynamic limit. 
However, these erratic oscillations are quite persistent in finite systems and
should dominate the dynamics of even fairly large, but finite populations for
many generations, see Figs.~\ref{oscill_diff_L} and \ref{L4096_fourier}.
These features, which appear to be more realistic from an ecological point of 
view than the regular (and initial-conditions dependent) oscillations predicted
by the conventional Lotka--Volterra equations, have been observed in other 
stochastic lattice predator--prey models as well 
\cite{Tome,Lip,Provata,TiborDroz,MGT1}, and likely represent a generic 
feature of such systems. 
This possibly has direct implications as it might shed further light on issues
of particular ecological and biological relevance, such as the emergence of 
(quasi-)oscillatory behavior and spontaneous pattern formation as results of 
stochastic fluctuations.

Typically, in two and three dimensions, when the fixed point is a focus (at 
large values of the predation rate $\lambda$), the species coexistence phase 
is characterized by the formation of complex and correlated patterns, as the 
result of the interaction and the propagation of the traveling wave fronts of 
predators and prey, which in turn cause the overall population oscillations 
\cite{Movies1}. 
The spatial structure of these patterns have been studied by computing the 
correlations functions, whilst their dynamical properties have been 
investigated through the computation of the functional dependence of the 
frequency of the resulting oscillations.
We have found that the typical frequency of the stochastic oscillations are
markedly reduced by fluctuation effects (i.e., compared to the mean-field
predictions).
 
Near their extinction threshold in the coexistence regime, the predators are 
largely localized in clusters interspersed in a sea of prey, with an active
reaction zone at their boundaries \cite{Movies1}. 
We have carefully analyzed the critical properties of the system by computing 
various critical exponents and have checked that the active to absorbing phase
transition belongs to the directed percolation (DP) universality class, with 
upper critical dimension $d_c=4$.  
Field-theoretic arguments support this conclusion: starting from the master
equation, we have constructed a field theory representation of the involved
stochastic processes, which near the predator extinction threshold can be 
mapped onto Reggeon field theory.
By utilizing tools of statistical mechanics (mean-field treatment together 
with field-theoretic and renormalization group arguments) we thus obtain a 
general qualitative understanding of the properties of the system and also a 
quantitative predictions of its behavior in the vicinity of the extinction 
threshold. 
In particular, the critical exponents governing the various statistical 
properties of the populations densities near the threshold are argued to be 
the same as in directed percolation (in dimensions $d>1$).
Also, active to absorbing state phase transitions and DP critical exponents 
were (numerically) reported in studies of other stochastic lattice 
predator--prey model variants \cite{Lip,TiborDroz,Boccara,Albano,MGT1}. 
Thus, our field-theoretic analysis should apply to a broader class of 
stochastic models than the ones considered here.

We have also discussed the one-dimensional case where, due to its special 
topology, the site occupation restriction on the lattice imply a `caging 
effect' resulting in species segregation and very slow coarsening of the 
predator domains in the system, which eventually evolves towards a lattice 
filled with prey.

Finally, we remark that in stark contrast with the deterministic 
Lotka--Volterra system, whose mathematical features are well-known to be 
quite unstable with respect to any model perturbations, the {\em stochastic}
spatial version is quite generic, and its overall features are rather
robust against model variations.
In particular, as noted also in Refs.~\cite{Lip,Provata}, we have checked that
the presence or absence of explicit particle diffusion does not qualitatively 
affect the properties of the system, since in our site-restricted system
species proliferation is generated by the offspring production processes.
Similarly, we have found that when the predation process is spatially biased,
near the extinction threshold only non-universal details change, but the DP
critical behavior still applies, and more generally, the overall picture  
drawn here remains valid in the active coexistence phase as well.
An intriguing situation is obtained when one considers a stochastic lattice 
predator-prey system with a next-nearest-neighbor (NNN) interaction among the
competing species, as well as a short-range exchange process \cite{MGT1}.
In this case a subtle interplay emerges between the NNN interaction and the
nearest-neighbor (NN) exchange or `mixing':
When the latter is `slow', due to the presence of correlations, this system 
also undergoes a DP-type phase transition (in dimensions $1 < d \leq 4$), as 
does the SLLVM studied in this work \cite{MGT1}.
However, when the value of the mixing rate is raised, the simple short-range 
exchange processes `wash out' the correlations and the system undergoes a 
{\em first-order phase transition} as in fact predicted by mean-field theory
\cite{MGT1}.
Whereas the rate equations predict entirely different behavior of the NNN 
system, which once more reflects the instability of the classical 
Lotka--Volterra model, it is quite remarkable that in the absence of explicit 
species mixing through particle exchange, the fluctuations render the 
properties of the NNN model akin to the simple SLLVM with only 
nearest-neighbor interactions.
In marked contrast with its mean-field counterpart, the stochastic lattice
Lotka--Volterra model is thus quite stable with respect to model 
modifications. 
On the other hand, as the mean-field regime is expected to be reached when the
exchange process allows the mixing of {\em all} the particles (and not only 
the immediate nearest neighboring ones) with an infinitely fast rate 
\cite{Durrett}, the fact that mean-field like behavior, characterized by a 
first-order phase transition, already appears unexpectedly even for finite NN 
exchange rates \cite{MGT1} is another quite intriguing feature of the NNN 
model. 
This is even more surprising since we have checked that fast diffusion affects
neither the critical nor the qualitative properties of the SLLVM studied in 
this paper.

\acknowledgments
This work was in part supported by U.S. National Science Foundation, through 
grants NSF DMR-0088451, DMR-0308548, and DMR-0414122.
MM acknowledges the support of the Swiss National Science Foundation through
Fellowship No. 81EL-68473, and of the German Alexander von Humboldt 
Foundation through Fellowship No. IV-SCZ/1119205 STP.
We would like to thank T. Antal, J. Banavar, E. Frey, P. L. Krapivsky, 
R. Kulkarni, T. Newman, G. Pruessner, B. Schmittmann, and R. K. P. Zia for 
inspiring and helpful discussions.

\appendix
\section{Kolmogorov's theorem and Bendixson--Dulac test}

In this appendix, we discuss the application of a general theorem due to 
Kolmogorov \cite{Kolmogorov} and of the so-called Bendixson--Dulac 
\cite{Sigmund} to the coupled rate equations (\ref{EOMinter}) and 
(\ref{EOMinter1}). 

We start with a theorem by Kolmogorov, who studied the mathematical properties
of two-species (mean-field type) rate equations of predator--prey models of 
the following form \cite{Kolmogorov,May1}:
\begin{eqnarray}
\label{KolmA}
  \dot{a}(t) &=& a(t) \, G(a(t),b(t)) \ , \\
\label{KolmB}
  \dot{b}(t) &=& b(t) \ F(a(t),b(t)) \ ,
\end{eqnarray}

Kolmogorov demonstrated that the generic system (\ref{KolmA}), (\ref{KolmB})
is characterized {\em either} by a stable fixed point {\em or} by a limit 
cycle, if $F$ and $G$ satisfy the following conditions \cite{May1}:
\begin{eqnarray}
\label{K1}
  \frac{\partial F}{\partial a} < 0 \; &;&  \; b \, 
  \frac{\partial F}{\partial b} + a \, \frac{\partial F}{\partial a} < 0 \ ; \\
\label{K3}
  b \, \frac{\partial G}{\partial b} + a \, \frac{\partial G}{\partial a} >
  0 \; &;& \; F(0,0) > 0 \ ;\\
\label{K5}
   \frac{\partial G}{\partial a} &<& 0.
\end{eqnarray}
In addition, there should exist three positive quantities, $k_i > 0$ 
($i=1,2,3$), such that $F(0,k_1) = 0$, $F(k_2,0) = 0$, $G(k_3,0) = 0$, and
$k_2 > k_3$.

We now consider the models studied Sec.~II.C and apply Kolmogorov's theorem
to these systems. 
In the case considered there, we specifically have: 
\begin{eqnarray}
F(a,b)&=&\sigma \left( 1 - \zeta^{-1} \, a - \rho^{-1} \, b \right) 
  - \lambda \, a, \\ 
  G(a,b) &=& \lambda \, b - \mu.
\end{eqnarray}
Thus, condition (\ref{K5}) is not fulfilled since 
$\partial G / \partial a = 0$. 
This means that Kolmogorov's theorem does not apply here (not even 
in the complete absence of growth-limiting terms, i.e., for
$\zeta = \rho = \infty$) and cannot ensure the existence of a stable fixed 
point or a limit cycle.

We now turn to the Bendixson--Dulac method which is a general approach to test 
whether a dynamical system of two two coupled differential equations admits 
periodic orbit solution. 
This method generally applies to any differential equation system 
${\dot{\bm x}} = {\bm f(\bm x)}$ of two variables ${\bm x} = (x_1, x_2)$, and 
states that there are no periodic orbits if ${\rm div} {\bm f(\bm x)} = 
\partial_{x_1} f_1(\bm x) + \partial_{x_2} f_2(\bm x) \neq 0$ and has only 
{\em one} sign in the whole space. 
As a consequence, for a strictly positive function ${\cal B}({\bm x})$, if 
${\rm div} \left({\cal B}({\bm x}){\bm f(\bm x)}\right) \neq 0$ and does not 
change its sign in the whole space, then ${\dot{\bm x}} = {\bm f(\bm x)}$ 
admits no periodic orbit \cite{nonlin}.
In addition, if ${\rm div} \left({\cal B}({\bm x}){\bm f(\bm x)}\right) = 0$, 
there exists a constant of motion for the original equation 
${\dot{\bm x}} = {\bm f(\bm x)}$ \cite{nonlin,Sigmund}.

Here, following the lines of Ref.~\cite{Sigmund} (Chap.~4), we apply the 
general Bendixson--Dulac method to Eqs.~(\ref{EOMinter}), (\ref{EOMinter1}) by
rewriting the latter as ${\dot a}=f_1 = a \, G(a,b)$ and 
${\dot b}=f_2 = b \, F(a,b)$.
We then consider a Dulac (auxiliary) function ${\cal B} = a^{\alpha-1} b^{-1}$ 
and apply the Bendixson--Dulac test. 
By computing the divergence of $({\cal B}f_1, {\cal B}f_2)$ and choosing  
$\alpha = \sigma / \rho \lambda$, one  finds ${\rm div} 
\left({\cal B}({\bm x}){\bm f(\bm x)}\right) = \partial_a 
  ({\cal B}f_1 ) + \partial_b ({\cal B}f_2) = - \mu \alpha {\cal B}$. 
Thus, according to the Bendixson--Dulac criterion \cite{nonlin,Sigmund} the 
existence of periodic orbits would require 
${\rm div} \left({\cal B}({\bm x}){\bm f(\bm x)}\right) = 0$.
Hence, for  Eqs.~(\ref{EOMinter}) and (\ref{EOMinter1}), periodic orbits are 
only possible when $\alpha = \sigma /\rho \lambda = 0$, i.e., for an infinite 
carrying capacities of prey, $\rho = \infty$.

In summary, Eqs.~(\ref{EOMinter}), (\ref{EOMinter1}) with finite (positive) 
rates $\mu$, $\sigma$, and $\lambda$, do {\em not} admit periodic orbits, 
except when $\rho = \infty$. 
In this special case, the Bendixson--Dulac method ensures that there exists a 
constant of motion  \cite{Sigmund}.


\begin{thebibliography}{99}

\bibitem{Lotka}
  A. J. Lotka, Proc. Natl. Acad. Sci. U.S.A. {\bf 6}, 410 (1920); 
  A. J. Lotka, J. Amer. Chem. Soc. {\bf 42}, 1595 (1920).

\bibitem{Volterra}
  V. Volterra, Mem. Accad. Lincei {\bf 2}, 31 (1926);
  V. Volterra, {\em Le\c cons sur la th\'eorie math\'ematique de la 
	lutte pour la vie} (Gauthiers-Villars, Paris 1931).

\bibitem{May1}
  {\em Theoretical Ecology}, edited by R. M. May 
	(Sinauer Associates, Sunderland, 1981); 
  {\em Population Regulation and Dynamics}, Proceedings of a Royal 
	Society discussion meeting, edited by M. P. Hassel and R. M. May 
	(London, The Royal Society, 1990); 
  R. M. May, {\em Stability and Complexity in Model Ecosystems},
  (Princeton University Press, Princeton, 1973).

\bibitem{Haken}
  H. Haken, {\em Synergetics} (Springer-Verlag, New York, 3rd ed. 1983).

\bibitem{Neal}
  D. Neal, {\em Introduction to Population Biology} 
  	(Cambridge University Press, Cambridge, 2004).

\bibitem{Maynard}
  J. Maynard Smith, {\em Models in Ecology} 
  	(Cambridge University Press, Cambridge, 1974).

\bibitem{Murray}
  J. D. Murray, {\em Mathematical Biology} Vols. I and II
	(Springer-Verlag, New York, 2002).

\bibitem{Kolmogorov}
  A. N. Kolmogorov, {\em Sulla Teoria di Volterra della Lotta per 
	l'Esistezza}, Giorn. Instituto Ital. Attuari, 7, 74-80 (1936).
	
\bibitem{Montroll}
  N. S. Goel, S. C. Maitra and E. W. Montroll, Rev. Mod. Phys. {\bf 43}, 
	231 (1971).

\bibitem{Picard}
	G. Picard and T. W. Watson, Phys. Rev. Lett. {\bf 48}, 1610 (1982).

\bibitem{Sigmund}  
  J. Hofbauer and K. Sigmund,	
	{\em Evolutionary Games and Population Dynamics} 
	(Cambridge University Press, Cambridge, 1998).

\bibitem{Leonard}
  R. M. May and W. Leonard, SIAM J. Appl. Math. {\bf 29}, 243 (1975).
	
\bibitem{Durrett}
  R. Durrett, SIAM Review {\bf 41}, 677 (1999).

\bibitem{data}
  C. Elton and M. Nicholson, J. Anim. Ecol. {\bf 11}, 215 (1942).

\bibitem{spatial}
  P. Rohani, R. M. May, and M. P. Hassell, 
	J. Theor. Biol. {\bf 181}, 97 (1996);
	{\em Modeling Spatiotemporal Dynamics in Ecology}, 
	edited by J. Bascompte and R. V. Sol\'e, (Springer, 1998).

\bibitem{Tome}
  J. E. Satulovsky and T. Tom\'e, Phys. Rev. E {\bf 49}, 5073 (1994).

\bibitem{Lip}
  A. Lipowski and D. Lipowska, Physica A {\bf 276}, 456 (2000);
  A. Lipowski, Phys. Rev. E {\bf 60}, 5179 (1999);
  M. Kowalick, A. Lipowski, and A. L. Ferreira, 
	{\em ibid.} {\bf 66}, 066107 (2002).

\bibitem{TiborDroz}
  T. Antal and M. Droz, Phys. Rev. E 63, 056119 (2001).

\bibitem{McKane}
  A. J. McKane and T. J. Newman, Phys. Rev. Lett. {\bf 94}, 218102 (2005).

\bibitem{Dunbar}
  S. R. Dunbar, J. Math. Biol. {\bf 17}, 11 (1983);
	Trans. Amer. Math. Soc. {\bf 268}, 557 (1984).

\bibitem{Provata}
  A. Provata, G. Nicolis, and F. Baras, 
	J. Chem. Phys. {\bf 110}, 8361 (1999);
  G. A. Tsekouras and A. Provata, Phys. Rev. E {\bf 65}, 016204 (2001).

\bibitem{Matsuda}
  H. Matsuda, N. Ogita, A. Sasaki, and K. Sat$\rm \overline{o}$,
  	Prog. Theor. Phys. {\bf 88}, 1035 (1992).

\bibitem{cyclic}
  L. Frachebourg and P. L. Krapivsky, J. Phys. A {\bf 31}, L287 (1998);
  L. Frachebourg, P. L. Krapivsky, and E. Ben-Naim, 
	Phys. Rev. E {\bf 54}, 6186 (1996);
  	Phys. Rev. Lett. {\bf 77}, 2125 (1996); 
  G. Szab\'o and T. C. Cz\'aran, Phys. Rev. E {\bf 63}, 061904 (2001);
  	{\em ibid.} {\bf 64}, 042902 (2002);
  G. Szab\'o and G. A. Sznaider, {\em ibid.} {\bf 69}, 031911 (2004).

\bibitem{Shnerb}
  E. Bettelheim, O. A. Nadav, and N. M. Shnerb,  
	Physica E {\bf 9}, 600 (2001); 
  N. M. Shnerb and O. Agam, {\tt e-print cond-mat/9903408}.

\bibitem{Boccara}
  N. Boccara, O. Roblin and M. Roger, Phys. Rev. E {\bf 50}, 4531 (1994).

\bibitem{Albano}
  A. F. Rozenfeld and E. V. Albano, Physica A {\bf 266}, 322 (1999);
  R. Monetti, A. F. Rozenfeld, and E. V. Albano, 
	Physica A {\bf 283}, 52 (2000);
  A. F. Rozenfeld and E. V. Albano, Phys. Rev. E {\bf 63}, 061907 (2001);
  A. F. Rozenfeld and E. V. Albano, Phys. Lett. A {\bf 332}, 361 (2004).
 
\bibitem{Droz}
  M. Droz and A. Pekalski, Phys. Rev. E {\bf 63}, 051909 (2001).

\bibitem{DP}
  W. Kinzel, in {\em Percolation structures and concepts}, 
	Annals of the Israel Physical Society {\bf 5}, 425 (1983);
  P. Grassberger, J. Phys. A {\bf 29}, 7013 (1996).

\bibitem{DPreviews}
  H. Hinrichsen, Adv. Phys. {\bf 49}, 815 (2000);
  H. K. Janssen and U. C. T\"auber, Ann. Phys. (NY) {\bf 315}, 147 (2005).

\bibitem{Janssen}
  H. K. Janssen, J. Stat. Phys. {\bf 103}, 801 (2001).

\bibitem{MGT1}
  M. Mobilia, I. T. Georgiev, and U. C. T\"auber, Phys. Rev. E {\bf 73}, 040903(R) (2006);
	{\tt e-print:q-bio.PE/0508043}.

\bibitem{nonlin}
  D. W. Jordan and P. Smith, {\em Nonlinear Ordinary Differential Equations},
  (Oxford University Press, Oxford, 3rd ed. 1999).
 
\bibitem{Riordan}
  J. Riordan, C. R. Doering, and D. ben-Avraham, Phys. Rev. Lett. {\bf 75}, 
	565 (1995).	

\bibitem{Levine}
  L. Pechenik and H. Levine, Phys. Rev. E {\bf 59}, 3893 (1999).

\bibitem{3S}
  Y. Fujii and M. Wadati, J. Phys. Soc. Jpn {\bf 66}, 3770 (1997);
  M. Mobilia and P.-A. Bares, Phys. Rev. E {\bf 63},  036121 (2001).	 

\bibitem{RD}
  {\em Nonequilibrium Statistical Mechanics in One Dimension}, 
	edited by V. Privman (Cambridge University Press, Cambridge, 1997);
  F. C. Alcaraz, M. Droz, M. Henkel, and V. Rittenberg, 
	Ann. Phys. (N.Y.) {\bf 230}, 250 (1994);
  M. Henkel, E. Orlandini, and J. Santos, 
	{\em ibid.} {\bf 259}, 163 (1997);
  D. C. Mattis and M. L. Glasser, Rev. Mod. Phys. {\bf 70}, 979 (1998).

\bibitem{DDS}
  B. Schmittmann abd R. K. P. Zia, in: 
	{\em Phase Transitions and Critical Phenomena}, Vol. 17, 
	edited by C. Domb and J. L. Lebowitz (Academic Press, New York, 1995).

\bibitem{Movies1}
  Two movies (`L256.mpg' and `L512.mpg') corresponding to the situation of 
  Fig.~\ref{2d_snap} (but with $\lambda = 2.1$ and lattice sizes 
  $256 \times 256$ and $512 \times 512$), as well as one movie 
  (`movie3.mpg') corresponding to the situation of Fig.~\ref{2d_snap_node} 
  (but with $\lambda = 0.18$  and lattice size $256 \times 256$) can be found 
  at {\tt http:// www.phys.vt.edu/$\sim$tauber/PredatorPrey/movies/}\, .

\bibitem{DPconjecture}
  H. K. Janssen, Z. Phys. B {\bf 42}, 151 (1981);
  P. Grassberger, Z. Phys. B {\bf 47}, 365 (1982).

\bibitem{Howard} 
  U. C. T\"auber, M. J. Howard, and B. P. Vollmayr-Lee, 
  J. Phys. A: Math. Gen. {\bf 38}, R79 (2005).
  
\bibitem{Odor}
  G. \'Odor and N. Menyh\'ard, Physica D {\bf 168}, 305 (2002).

\bibitem{DHT}
  O. Deloubri\`ere, H. J. Hilhorst, and U. C. T\"auber,
  Phys. Rev. Lett. {\bf 89}, 250601 (2002);
  H. J. Hilhorst, O. Deloubri\`ere, M. J. Washenberger, and  U. C. T\"auber,
  J. Phys. A: Math. Gen. {\bf 37}, 7063 (2004).

\bibitem{fred}
  F.~van~Wijland, Phys. Rev. E {\bf 63}, 022101 (2001). 
  
\end{thebibliography}
\end{document}